\documentclass[nofootinbib,floatfix,amsmath,amssymb,showpacs,superscriptaddress,notitlepage]{revtex4-1}
\usepackage{dcolumn}
\usepackage{multirow}
\usepackage[export]{adjustbox}
\usepackage[pdftex,bookmarks=true]{hyperref}
\hypersetup{colorlinks, citecolor=blue, filecolor=blue, linkcolor=blue, urlcolor=blue}
\newcommand{\slp}{\Sigma_c \rightarrow\Lambda_c \pi}
\newcommand{\lsp}{\Lambda_c \rightarrow \Sigma_c \pi}
\newcommand{\slpc}{\Sigma_c \Lambda_c \pi}
\newcommand{\lspc}{\Lambda_c \Sigma_c \pi}

\allowdisplaybreaks

\begin{document}

\title{$\Lambda_c\Sigma_c\pi$ coupling and $\Sigma_c \rightarrow\Lambda_c \pi$ decay in lattice QCD}

\author{K. U. Can}
\affiliation{Department of Physics, H-27, Tokyo Institute of Technology, Meguro, Tokyo 152-8551 Japan}
\author{G. Erkol}
\affiliation{Department of Natural and Mathematical Sciences, Faculty of Engineering, Ozyegin University, Nisantepe Mah. Orman Sok. No:34-36, Alemdag 34794 Cekmekoy, Istanbul Turkey}
\author{M. Oka}%
\affiliation{Department of Physics, H-27, Tokyo Institute of Technology, Meguro, Tokyo 152-8551 Japan}
\affiliation{Advanced Science Research Center, Japan Atomic Energy Agency, Tokai, Ibaraki, 319-1195 Japan}
\author{T. T. Takahashi}%
\affiliation{Gunma National College of Technology, Maebashi, Gunma 371-8530 Japan}

\date{\today}

\begin{abstract}
We evaluate the $\Lambda_c\Sigma_c\pi$ coupling constant ($G_{\lspc}$) and the width of the strong decay $\Sigma_c \rightarrow\Lambda_c \pi$ in 2+1 flavor lattice QCD on four different ensembles with pion masses ranging from 700 MeV to 300 MeV. We find $G_{\lspc}=18.332(1.476)_{\rm{stat.}}(2.171)_{\rm{syst.}}$ and the decay width $\Gamma(\Sigma_c \rightarrow\Lambda_c \pi)=1.65(28)_{\rm{stat.}}(30)_{\rm{syst.}}$~MeV on the physical quark-mass point, which is in agreement with the recent experimental determination.

\end{abstract}
\pacs{14.20.Lq, 12.38.Gc, 13.40.Gp }
\keywords{charmed baryons, pion couplings, lattice QCD}
\maketitle

\section{Introduction}
We have seen an immense progress on the physics of charmed baryons in the last decade and all the ground-state single-charmed baryons and several excited states, as predicted by the quark model, have been experimentally measured~\cite{Agashe:2014kda}. The properties of $\Sigma_c$ and $\Lambda_c$ baryons and the $\slp$ decay have been experimentally determined by E791~\cite{Aitala:1996cy}, FOCUS~\cite{Link:2000qs, Link:2001ee}, CLEO~\cite{Artuso:2001us, Athar:2004ni}, BABAR~\cite{Aubert:2008ax} and CDF~\cite{Aaltonen:2011sf} Collaborations. The world averages for $\Sigma_c$ and $\Lambda_c$ masses are $m_{\Sigma^{++}_c}=2453.97 \pm 0.14$ MeV and $m_{\Lambda^{+}_c}=2286.46 \pm 0.14$ MeV~\cite{Agashe:2014kda}. The $\Sigma_c$ has a width of $\Gamma_{\Sigma^{++}_c}=1.89^{+0.09}_{-0.18}$ MeV where it dominantly decays via strong $\Sigma_c\rightarrow \Lambda_c\pi$ channel. The strong decay $\slp$ has been studied in Heavy Hadron Chiral Perturbation Theory~\cite{Yan:1992gz, Huang:1995ke, Cheng:2015naa}, Light-front Quark Model~\cite{Tawfiq:1998nk}, Relativistic Quark Model~\cite{Ivanov:1999bk}, nonrelativistic Quark Model~\cite{Albertus:2005zy, Nagahiro:2016nsx}, $^3P_0$ Model~\cite{Chen:2007xf} and QCD Sum Rules~\cite{PhysRevD.79.056002}. Most recently, Belle Collaboration has measured the decay width of $\Sigma_c(2455)^{++}$ as $\Gamma=1.84\pm 0.04^{+0.07}_{-0.20}$~MeV and that of $\Sigma_c(2455)^{0}$ as $\Gamma=1.76\pm 0.04^{+0.09}_{-0.21}$~MeV~\cite{Lee:2014htd}.

We have recently extracted the electromagnetic form factors of baryons in lattice QCD~\cite{Can:2013zpa, Can:2013tna, Can:2015exa}. Motivated by the recent experimental measurements, in this work we broaden our program to include pion couplings of baryons. As a first step we evaluate the strong coupling constant $\Lambda_c\Sigma_c\pi$ and the width of the strong decay $\slp$ in 2+1 flavor lattice QCD. Our aim is to utilize this calculation as a benchmark for future calculations. This work is reminiscent of Refs.~\cite{Alexandrou:2007zz, Erkol:2008yj} where pion--octet-baryon coupling constants have been calculated in lattice QCD. 

Our work is organized as follows: In Section~\ref{sec:tfls} we present the theoretical formalism of our calculations of the form factors together with the lattice techniques we have employed to extract them. In Section~\ref{sec:resdis} we present and discuss our numerical results. Section~\ref{sec:sum} contains a summary of our findings.

\section{Theoretical formulation and lattice simulations}
\label{sec:tfls}
We begin with formulating the baryon matrix elements of the pseudoscalar current, which we evaluate on the lattice to compute the pion coupling constants. The pion has a direct coupling to the axial-vector current $A_\mu^a(x)=\bar{\psi}(x)\gamma_\mu\gamma_5 \frac{\tau^a}{2}\psi(x)$ as
\begin{equation}
	\langle 0 | A^a_\mu (0)| \pi^b(q)\rangle= i f_\pi q_\mu \delta^{ab}, \qquad a,b=1,2,3
\end{equation}
where $f_\pi=92$~MeV is the pion decay constant. Taking the divergence of the axial-vector current, we find the partially conserved axial-vector current (PCAC) hypothesis
\begin{equation}
	\partial^\mu A_\mu^a=f_\pi m_\pi^2\phi^a,
\end{equation}
where $\phi^a$ is the pion field with the normalization $\langle 0|\phi^a(0)|\pi^b(q)\rangle=\delta^{ab}$. The matrix element of the PCAC hypothesis between baryon states yields
\begin{align}\label{mpcac}
	\begin{split}
		\langle \mathcal{B^\prime}(p') | \partial^\mu A^3_\mu| \mathcal{B}(p)\rangle&= f_\pi m_\pi^2 \langle \mathcal{B}(p')|\phi^3(0)|\mathcal{B}(p)\rangle\\
		&=\left(\frac{M_\mathcal{B}M_\mathcal{B}^\prime}{E\,E^\prime}\right)^{1/2}\frac{f_\pi m_\pi^2}{m_\pi^2-q^2} G_{\mathcal{B}^\prime \mathcal{B} \pi}(q^2) \bar{u}_{\mathcal{B}^\prime}(p^\prime)i\gamma_5 \frac{\tau^3}{2} u_\mathcal{B}(p).
	\end{split}
\end{align}
Here, $u_B$ is the baryon Dirac spinor, $\mathcal{B}$ ($\mathcal{B}^\prime$) denotes the incoming (outgoing) baryon and $M_\mathcal{B}$ ($M_\mathcal{B}^\prime$), $E$ ($E^\prime$) and $p$ ($p^\prime$) are the rest mass, energy and the four momentum of the baryon, respectively. We specifically consider the axial isovector current $A^3_\mu$ and the pion field $\phi^3$ with momentum $q=p^\prime-p$. $G_{\mathcal{B}^\prime \mathcal{B} \pi}$ is the $\mathcal{B}^\prime \mathcal{B} \pi$ coupling constant.

At the quark level we have the axial Ward-Takahashi identity
\begin{equation}\label{wti}
	\partial^\mu A_\mu^a=2 m_q P^a,
\end{equation}
where $P^a(x)=\bar{\psi}(x)\gamma_5 \frac{\tau^a}{2}\psi(x)$ is the pseudoscalar current and $\psi(x)$ is the isospin doublet quark field. Inserting Eq.~\eqref{wti} into Eq.~\eqref{mpcac}, we find the baryon-baryon matrix elements of the pseudoscalar current
\begin{equation}
	\label{bme}
	\langle \mathcal{B^\prime}(p') | P^3| \mathcal{B}(p)\rangle=\left(\frac{M_\mathcal{B}M_\mathcal{B}^\prime}{E\,E^\prime}\right)^{1/2} \frac{f_\pi m_\pi^2}{m_\pi^2-q^2} \frac{G_{\mathcal{B}^\prime \mathcal{B} \pi}(q^2)}{2m_q}  \bar{u}_{\mathcal{B}^\prime}(p^\prime)i\frac{\tau^3}{2} \gamma_5 u_\mathcal{B}(p),
\end{equation}
which we use to extract $G_{\mathcal{B}^\prime \mathcal{B} \pi}$. We use the PACS-CS determined values~\cite{Aoki:2008sm} for pion decay constant, $f_\pi$, pion mass, $m_\pi$, and the quark mass, $m_q$, on each ensemble. 

While the matrix element in Eq.~\eqref{bme} is derived by a PCAC prescription we can extract the pseudoscalar matrix elements on the lattice directly by using the following ratio
\begin{align}
\begin{split}\label{ratio}
	&R(t_2,t_1;{\bf p}^\prime,{\bf p};\mathbf{\Gamma};\mu)=
	\cfrac{\langle G^{{\cal B^\prime P B}}(t_2,t_1; {\bf p}^\prime, {\bf p};\mathbf{\Gamma})\rangle}{\langle G^{{\cal B^\prime}{\cal B^\prime}}(t_2; {\bf p}^\prime;\Gamma_4)\rangle}\left[\frac{\langle G^{{\cal B B}}(t_2-t_1; {\bf p};\Gamma_4)\rangle \langle G^{{\cal B^\prime B^\prime}}(t_1; {\bf p}^\prime;\Gamma_4)\rangle \langle G^{{\cal B^\prime B^\prime}}(t_2; {\bf p}^\prime;\Gamma_4)\rangle}{\langle G^{{\cal B^\prime B^\prime}}(t_2-t_1; {\bf p}^\prime;\Gamma_4)\rangle\langle G^{{\cal B B}}(t_1; {\bf p};\Gamma_4)\rangle \langle G^{{\cal B B}}(t_2; {\bf p};\Gamma_4)\rangle} \right]^{1/2},
\end{split}
\end{align}
where the baryonic two-point and three-point correlation functions are respectively defined as
\begin{align}
	\begin{split}\label{twopcf}
	&\langle G^{{\cal BB}}(t; {\bf p};\Gamma_4)\rangle=\sum_{\bf x}e^{-i{\bf p}\cdot {\bf x}}\Gamma_4^{\alpha\beta} \times \langle \text{vac} | T [\eta_{\cal B}^\alpha(\mathbf{x},t) \bar{\eta}_{{\cal B}}^{\beta}(\mathbf{0},0)] | \text{vac}\rangle,
	\end{split}\\
	\begin{split}\label{thrpcf}
	&\langle G^{{\cal B^\prime P B}}(t_2,t_1; {\bf p}^\prime, {\bf p};\mathbf{\Gamma})\rangle=-i\sum_{{\bf x_2},{\bf x_1}} e^{-i{\bf p^\prime}\cdot {\bf x_2}} e^{i({\bf p^\prime-p})\cdot {\bf x_1}} \Gamma^{\alpha\beta} \langle \text{vac} | T [\eta_{\cal B^\prime}^{\alpha}(\mathbf{x_2},t_2) P^3 (\mathbf{x_1},t_1) \bar{\eta}_{{\cal B}}^{\beta}(\mathbf{0},0)] | \text{vac}\rangle,
	\end{split}
\end{align}
with $\Gamma_i=\gamma_i\gamma_5\Gamma_4$ and $\Gamma_4\equiv (1+\gamma_4)/2$. $t_1$ is the time when the external pseudoscalar field interacts with a quark and $t_2$ is the time when the final baryon state is annihilated. 

The baryon interpolating fields are chosen as
\begin{align}
		\eta_{\Sigma_c}(x)&=\epsilon^{ijk}\left\{[u^{T i}(x) C \gamma_5 c^j(x)]d^k(x)+[d^{T i}(x) C \gamma_5 c^j(x)]u^k(x)\right\},\\
		\eta_{\Lambda_c}(x)&=\epsilon^{ijk}\left\{[2 u^{T i}(x) C \gamma_5 d^j(x)]c^k(x)+[u^{T i}(x) C \gamma_5 c^j(x)]d^k(x)-[d^{T i}(x) C \gamma_5 c^j(x)]u^k(x)\right\},
\end{align}
where $i$, $j$, $k$ denote the color indices and $C=\gamma_4\gamma_2$. In the large Euclidean time limit, $t_2-t_1$ and $t_1\gg a$, the ratio in Eq.~(\ref{ratio}) reduces to the desired form
\begin{equation}\label{desratio}
	R(t_2,t_1;{\bf p^\prime},{\bf p};\Gamma;\mu)\xrightarrow[t_2-t_1\gg a]{t_1\gg a} \Pi({\bf p^\prime},{\bf p};\Gamma;\gamma_5)=\left(\frac{1}{2E(E+M_{\cal B})}\right)^{1/2} \frac{q_k}{2m_\mathcal{B}} \frac{f_\pi m_\pi^2}{2m_q (m_\pi^2+Q^2)} G_{\mathcal{B}^\prime \mathcal{B} \pi}(q^2),
\end{equation}
where $Q^2=-q^2$. We measure the $\Lambda_c\Sigma_c\pi$ coupling constant for both kinematical cases with $\mathcal{B}^\prime=\Sigma_c$, $\mathcal{B}=\Lambda_c$ (denoted by $G_{\Sigma_c \Lambda_c \pi}$) and $\mathcal{B}^\prime=\Lambda_c$, $\mathcal{B}=\Sigma_c$ (denoted by $G_{\Lambda_c \Sigma_c \pi}$).

Here we summarize our lattice setup and refer the reader to Ref.~\cite{Can:2012tx} for the details since we employ the same setup in this work. We have run our lattice simulations on $32^3\times 64$ lattices with 2+1 flavors of dynamical quarks using the gauge configurations generated by the PACS-CS collaboration~\cite{Aoki:2008sm} with the nonperturbatively $\mathcal{O}(a)$-improved Wilson quark action and the Iwasaki gauge action. We use the gauge configurations at $\beta=1.90$ with the clover coefficient $c_{SW}=1.715$ having a lattice spacing of $a=0.0907(13)$ fm ($a^{-1}=2.176(31)$~GeV). We consider four different hopping parameters for the sea and the $u$,$d$ valence quarks, $\kappa_{sea},\kappa_{val}^{u,d}=$ 0.13700, 0.13727, 0.13754 and 0.13770, which correspond to pion masses of $\sim$ 700, 570, 410, and 300~MeV, respectively. 

We use the \emph{wall method} which does not require to fix sink operators in advance and hence allowing us to compute all baryon channels we are interested in simultaneously. However, since the wall sink/source is a gauge-dependent object, we have to fix the gauge, which we choose to be Coulomb. We extract the baryon masses from the two-point correlator with shell source and point sink, and use the dispersion relation to calculate the energy at each momentum transfer. 

Similar to our simulations in Ref.~\cite{Can:2012tx}, we choose to employ Clover action for the charm quark. While the Clover action is subject to discretization errors of $\mathcal{O}(m_q\,a)$, it has been shown that the calculations which are insensitive to a change of charm-quark mass are less severely affected by these errors~\cite{Bali:2011rd, Can:2012tx, Can:2013zpa, Can:2013tna, Can:2015exa}. Note that the Clover action we are employing here is a special case of the Fermilab heavy-quark action with $c_{SW}=c_E=c_B$~\cite{Burch:2009az}. We determine the hopping parameter of the charm quark nonperturbatively as $\kappa_{c}=0.1246$ by tuning the spin-averaged static masses of charmonium and heavy-light mesons to their experimental values~\cite{Can:2013tna}. 

We employ smeared source and wall sink which are separated by 12 lattice units in the temporal direction. Light and charm quark source operators are smeared in a gauge-invariant manner with the root mean square radius of $\langle r_l \rangle \sim 0.5$~fm and $\langle r_c \rangle=\langle r_l \rangle/3$ respectively. All the statistical errors are estimated via the jackknife analysis. In this work, we consider only the connected diagrams since the $P^3$ current is an isovector current and the relevant light quark disconnected diagrams vanish. 

We make our measurements on 100, 100, 200 and 315 configurations, respectively for each quark mass. In order to increase the statistics we take several different source points using the translational invariance along the temporal direction. We make momentum insertions in all directions and average over equivalent (positive and negative) momenta. Computations are performed using a modified version of Chroma software system~\cite{Edwards:2004sx} on CPU clusters along with QUDA~\cite{Babich:2011np,Clark:2009wm} for propagator inversion on GPUs.

\section{Results and discussion}
\label{sec:resdis}
Masses of the baryons in question are input parameters for form factor calculations. In Table~\ref{bar_mass}, we give $\Lambda_c$ and $\Sigma_c$ masses for four light-quark hopping-parameter values corresponding to each light-quark mass we consider. We extrapolate the masses to the physical point by a HH$\chi$PT procedure as outlined in Ref~\cite{PhysRevD.81.094505}. Our results are compared to those reported by PACS-CS~\cite{Namekawa:2013vu}, ETMC~\cite{Alexandrou:2014sha}, Briceno \emph{et al.}~\cite{Briceno:2012wt} and Brown \emph{et al.}~\cite{Brown:2014ena} and to the experimental values~\cite{Agashe:2014kda} in Table~\ref{bar_mass}.
%%%%%%%%%%%%%%%%%%%%%%%%%% Table - Baryon masses %%%%%%%%%%%%%%%%%%%%%%%%%%%
\begin{table}[thb]
	\caption{ We give $\Lambda_c$ and $\Sigma_c$ masses for four light-quark hopping parameter values corresponding to each light-quark mass we consider. For comparison we also give our extrapolated values of masses, together with those reported by other collaborations and the experimental values~\cite{Agashe:2014kda}. Quoted errors for other lattice works are combined errors from statistical, chiral and continuum extrapolations where available.}
	\label{bar_mass}
	\centering
	\setlength{\extrarowheight}{7pt}
	\begin{tabular*}{0.8\textwidth}{@{\extracolsep{\fill}}c|cccccc}
		\hline\hline 
		   $\kappa_{val}^{u,d}$ & 0.13700 & 0.13727 & 0.13754 & 0.13770 & &  \\
		\hline
		$M_{\Lambda_c}$~[GeV]  & 2.713(16) & 2.581(21) & 2.473(15) & 2.445(13) &   \\
		\hline\hline 
		 chiral point  & This work & PACS-CS~\cite{Namekawa:2013vu} &  ETMC~\cite{Alexandrou:2014sha} & Briceno \emph{et al.}~\cite{Briceno:2012wt} & Brown \emph{et al.}~\cite{Brown:2014ena} & Exp.~\cite{Agashe:2014kda}   \\
		\hline
		 $M_{\Lambda_c}$~[GeV]  & 2.412(15) & 2.333(122) & 2.286(27) & 2.291(66) & 2.254(79) & 2.28646(14)\\
		\hline 			
	\end{tabular*}
	\vspace{0.11in}
	\begin{tabular*}{0.8\textwidth}{@{\extracolsep{\fill}}c|cccccc}
		\hline 
		   $\kappa_{val}^{u,d}$ & 0.13700 & 0.13727 & 0.13754 & 0.13770 & &  \\
		\hline
		$M_{\Sigma_c}$~[GeV]  &  2.806(19) & 2.716(20) & 2.634(16) & 2.590(19) &   \\
		\hline 
		 chiral point  & This work & PACS-CS~\cite{Namekawa:2013vu} &  ETMC~\cite{Alexandrou:2014sha} & Briceno \emph{et al.}~\cite{Briceno:2012wt} & Brown \emph{et al.}~\cite{Brown:2014ena} & Exp.~\cite{Agashe:2014kda}   \\
		\hline
		 $M_{\Sigma_c}$~[GeV]  & 2.549(72) & 2.467(50) & 2.460(46) & 2.481(46) & 2.474(66) & 2.45397(14)\\
		\hline \hline			
	\end{tabular*}
\end{table}
%%%%%%%%%%%%%%%%%%%%%%%%%%%%%%%%%%%%%%%%%%%%%%%%%%%%%%%%%%%%%%%%%%

Our results lie $\sim 100$ MeV above the experimental values. This is due to our choice of $\kappa_c$ which we have tuned according to the spin-averaged static masses of the charmonium and open charm mesons in Ref.~\cite{Can:2013tna}. Although meson masses are in very good agreement with their experimental values, baryon masses are overestimated by $\sim 50-150$ MeV. Since the baryon masses only appear in kinematical terms in form factor calculations the sensitivity of the final results to mass deviations are negligible. A mistuning would effect other charm related observables however the quantities we extract in this work have no direct relation to the charm quark since it acts as a spectator quark. We have, on the other hand, confirmed in our previous works~\cite{Can:2013tna,Can:2013zpa,Can:2015exa} that charm quark observables are affected by less than $2\%$ by changing the $\kappa_c$ so that the mass deviates by $\sim 100$ MeV. We expect the effects, if any, to be similar or less in this work as well.

We make our analysis by considering two different kinematic cases where we choose the source particle as a $\Sigma_c$ or a $\Lambda_c$ particle. The first case corresponds to the $\slp$ transition where the particle at sink, that is $\Lambda_c$, is at rest since its momentum is projected to zero due to wall smearing. The second case is the $\lsp$ transition where $\Sigma_c$ is located at the sink point. A common practice to extract the form factors is to identify the regions where the ratio in Eq.~\ref{ratio} remains constant, namely forms a plateau with respect to the current-insertion time, $t_1$. However, due to a finite source-sink seperation, it might not always be possible to identify a clean plateau signal and an asymmetric (Gaussian smeared) source-(wall smeared) sink pair, as employed here would further affect the signal since different smearing procedures are known to cause different ground-state approaches. An ill-defined plateau range would be prone to excited state contamination which would introduce an uncontrolled systematic error. In order to check that our plateau analysis yields reliable results we compare the form factor values extracted by the plateau method to the ones extracted by a phenomenological form given as,
\begin{equation}
	\label{epf}
	R(t_2,t_1) = G_{\mathcal{B}^\prime \mathcal{B} \pi} + b_1 e^{- \Delta_1 t_1} + b_2 e^{- \Delta_2 (t_2-t_1)},
\end{equation}
where the first term is the form factor value we wish to extract and the coefficients $b_1$, $b_2$ and the mass gaps $\Delta_1$, $\Delta_2$ are regarded as free parameters.     
%%%%%%%%%%%%%%%%%%%%%% Figure - Plateau-EPF %%%%%%%%%%%%%%%%%
\begin{figure}[ht]
	\centering
	\includegraphics[width=.49\textwidth]{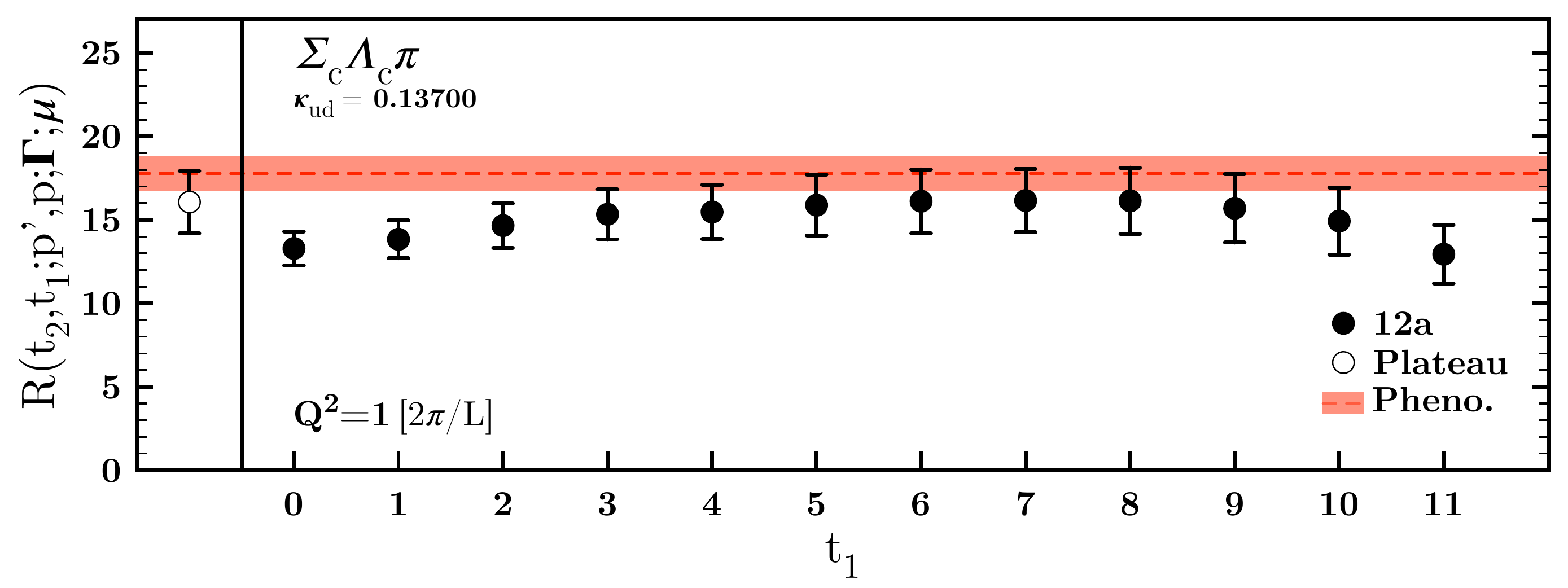}	\includegraphics[width=.49\textwidth]{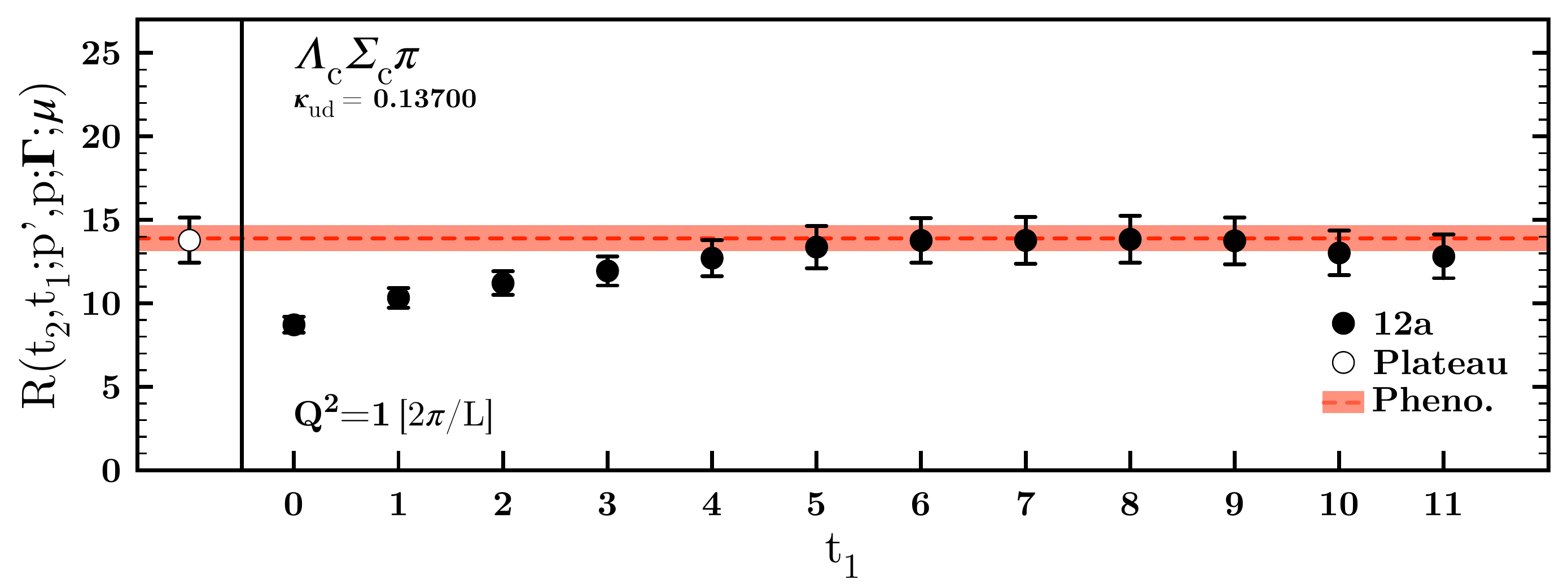}	\\
	\includegraphics[width=.49\textwidth]{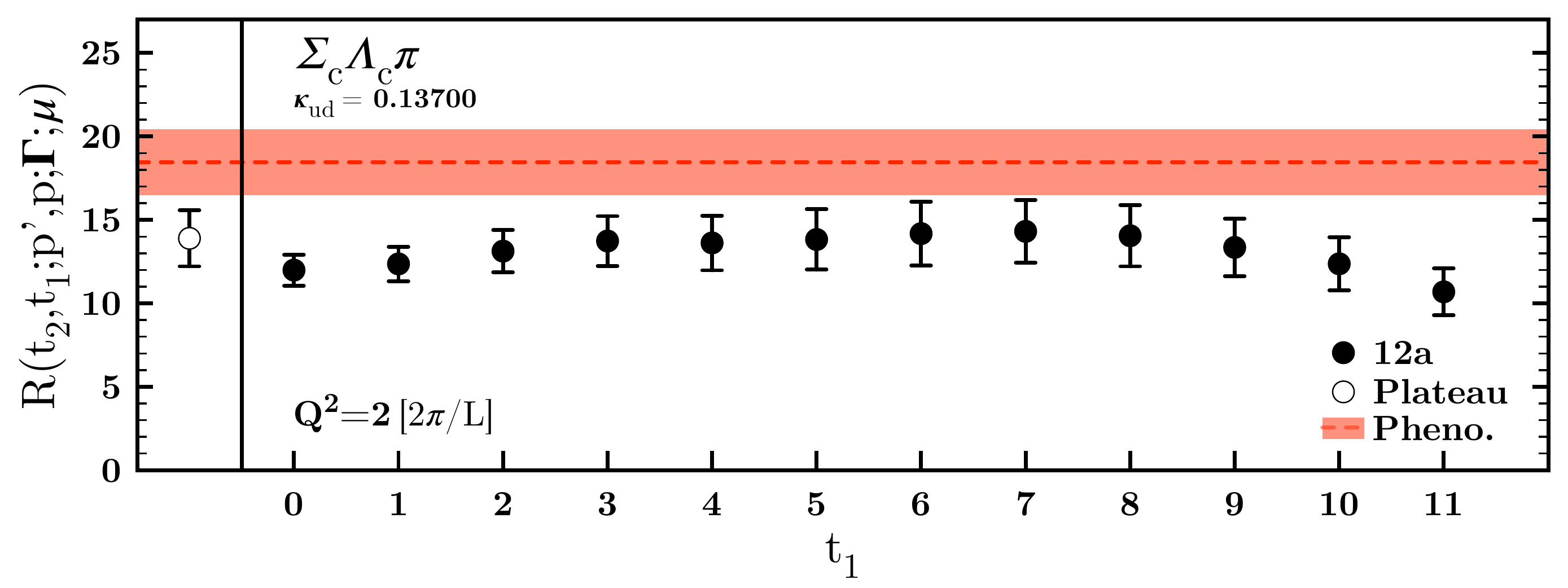}	\includegraphics[width=.49\textwidth]{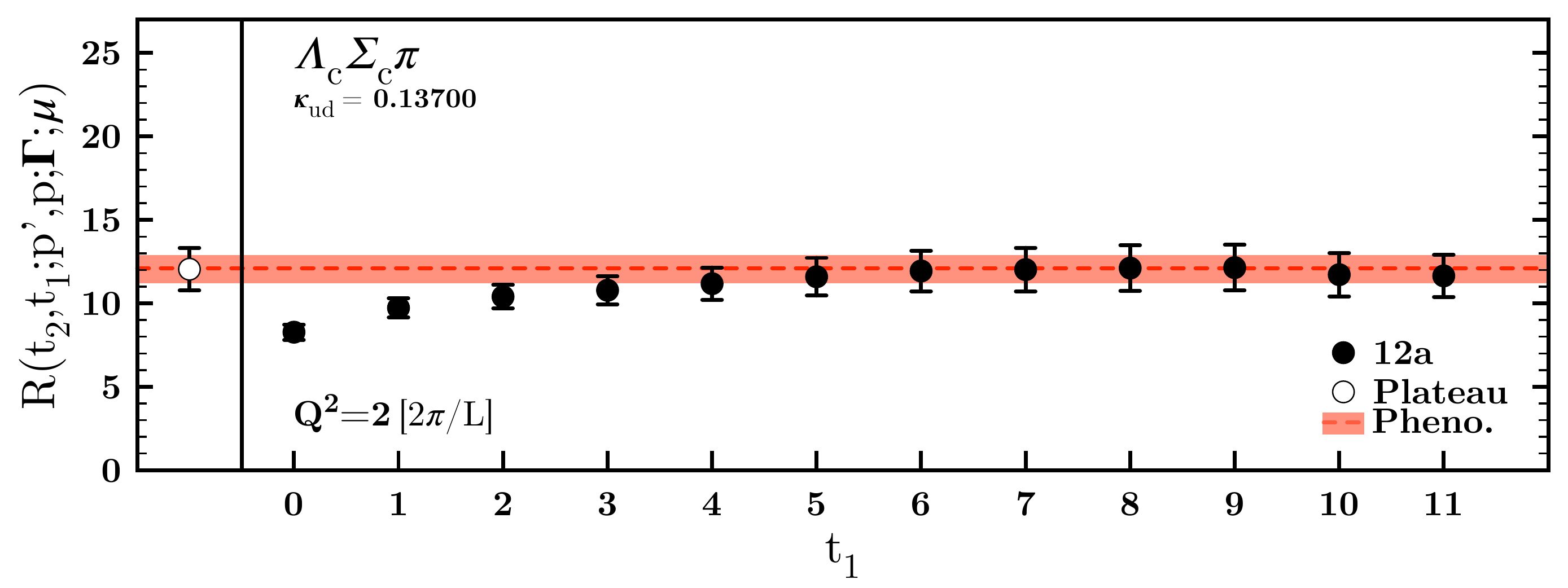}	\\
	\includegraphics[width=.49\textwidth]{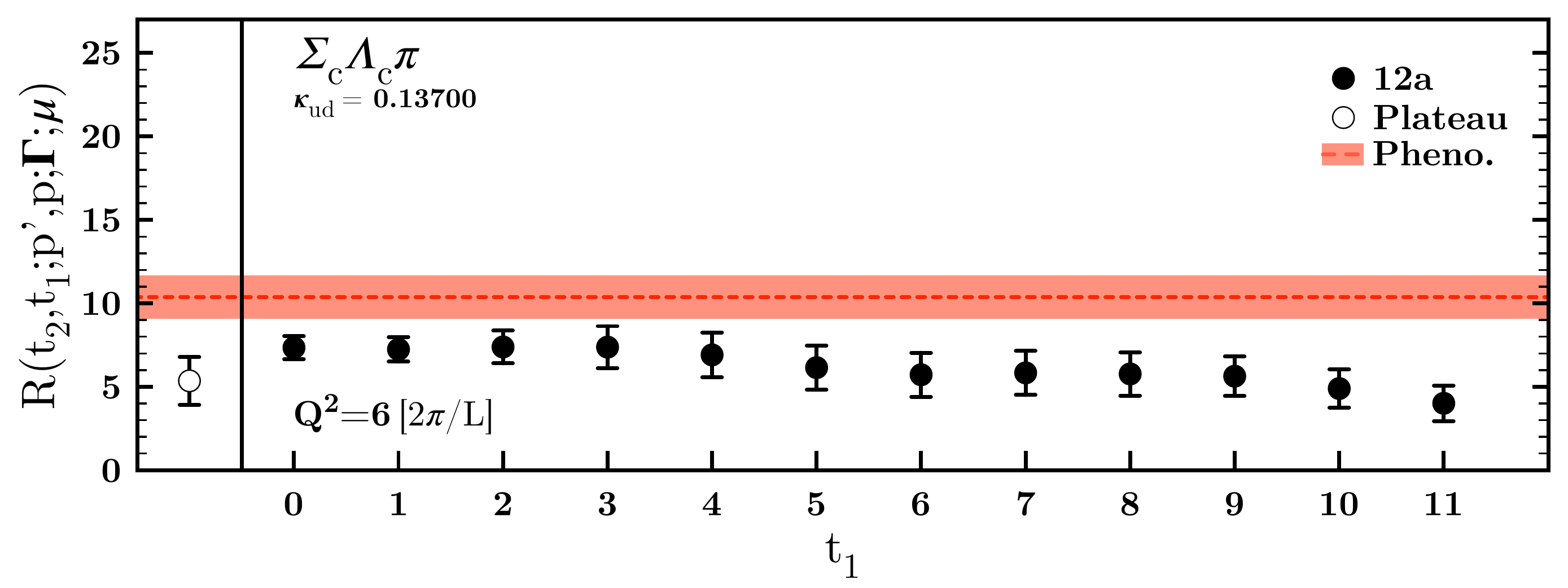}	\includegraphics[width=.49\textwidth]{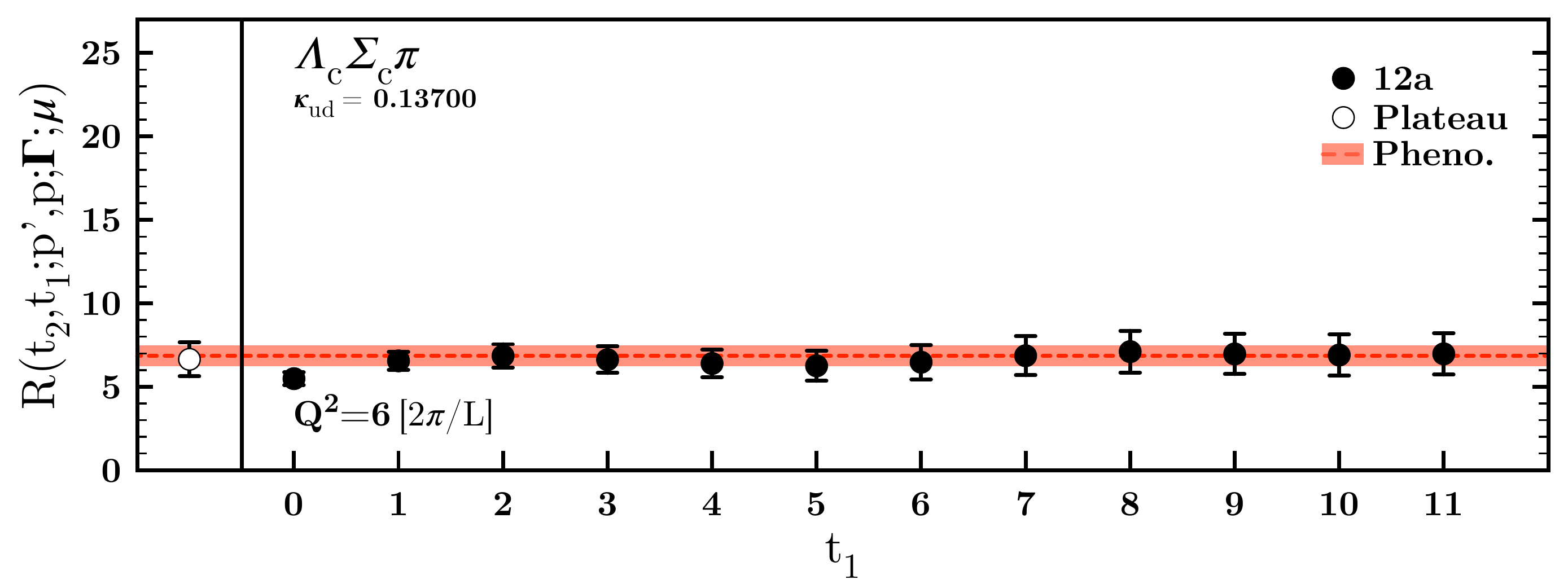}				
	\caption{A comparison of plateau fit to phenomenological form fit illustrated on the heaviest quark mass ensemble $\kappa^{u,d}=$0.13700 for the $\lsp$ (left) and $\slp$ (right) kinematical cases. Open symbols on the left panels indicate the best fit value to the identified plateau region. Red bands show the extracted value by a phenomenological form fit.  }
	\label{fig:plepf}
\end{figure}
% %%%%%%%%%%%%%%%%%%%%%%%%%%%%%%%%%%%%%%%%%%%%%%%%
%%%%%%%%%%%%%%%%%%%%%% Figure - 1214 %%%%%%%%%%%%%%%%%%%%
\begin{figure}[ht]
	\centering
	\includegraphics[width=.49\textwidth]{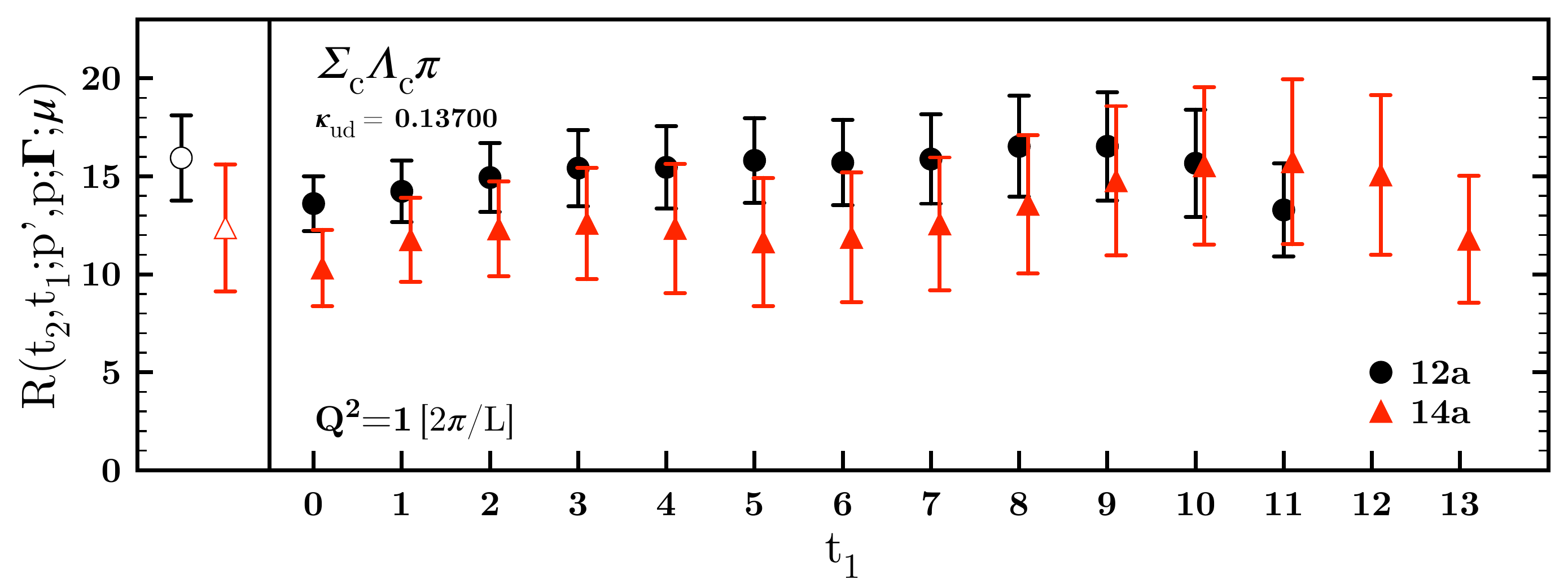}	\includegraphics[width=.49\textwidth]{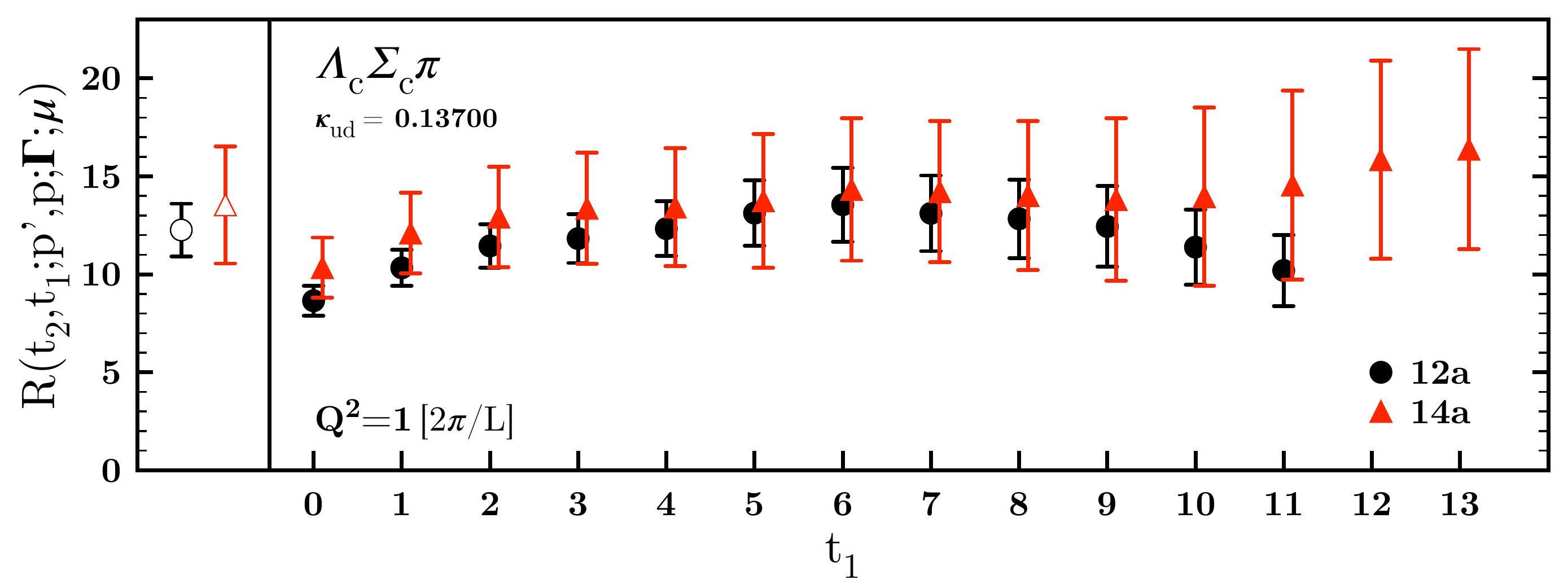}	\\
	\includegraphics[width=.49\textwidth]{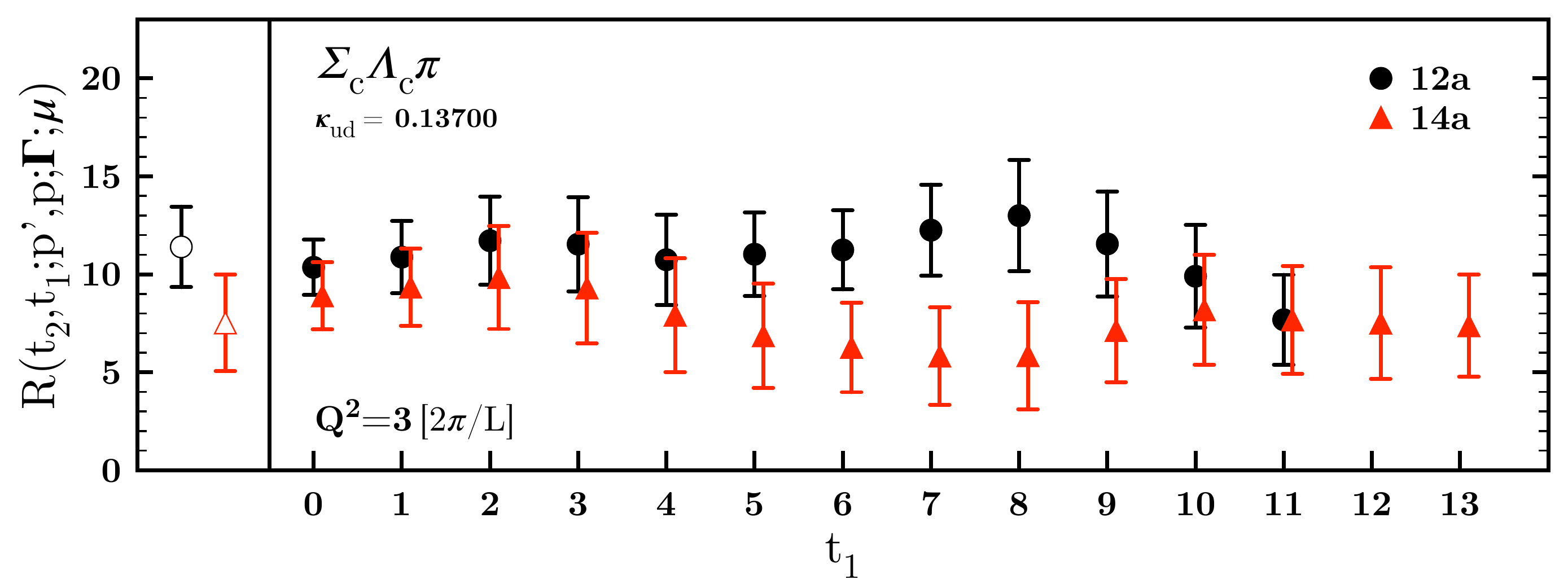}	\includegraphics[width=.49\textwidth]{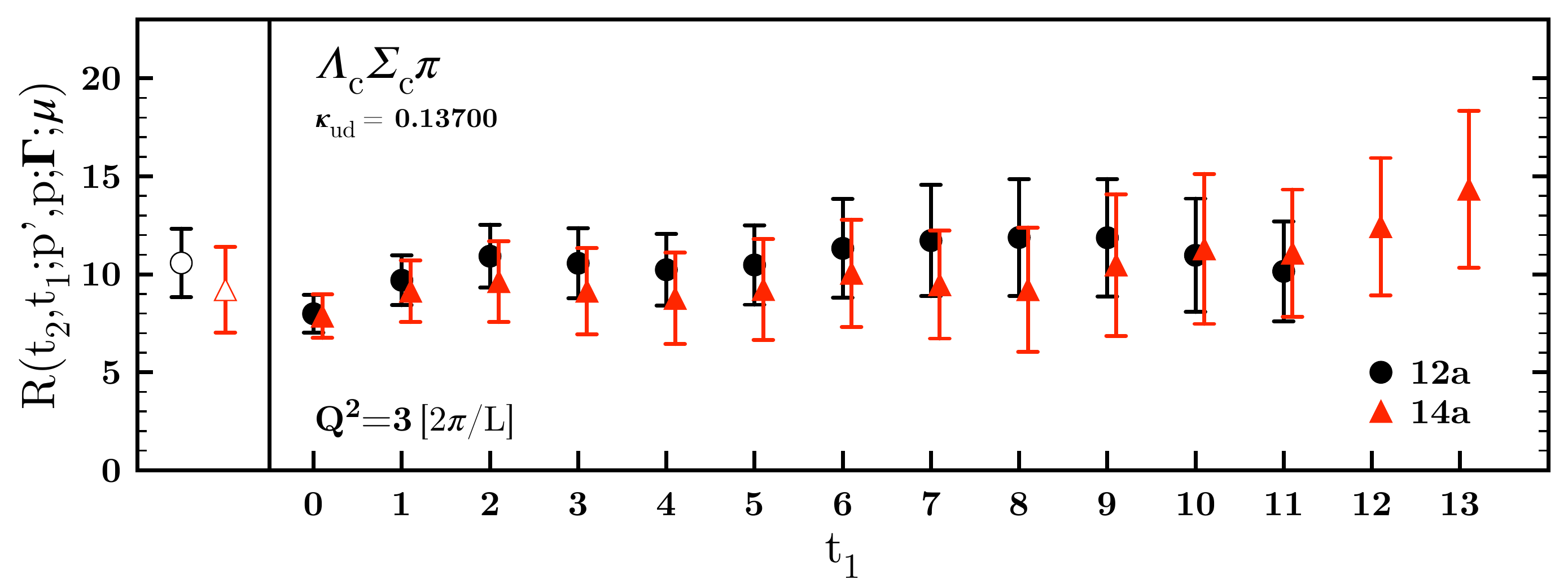}	\\
	\includegraphics[width=.49\textwidth]{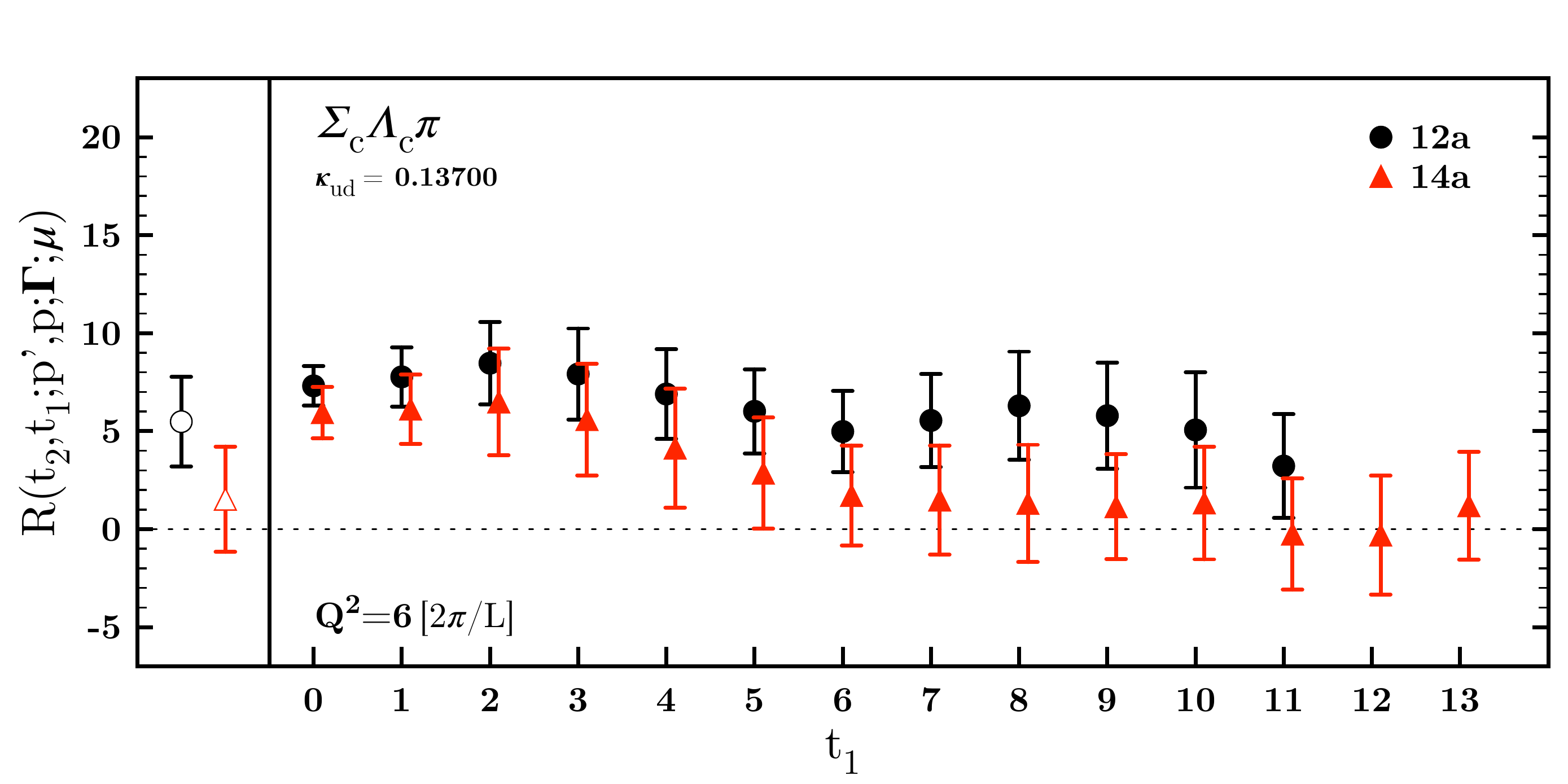}	\includegraphics[width=.49\textwidth]{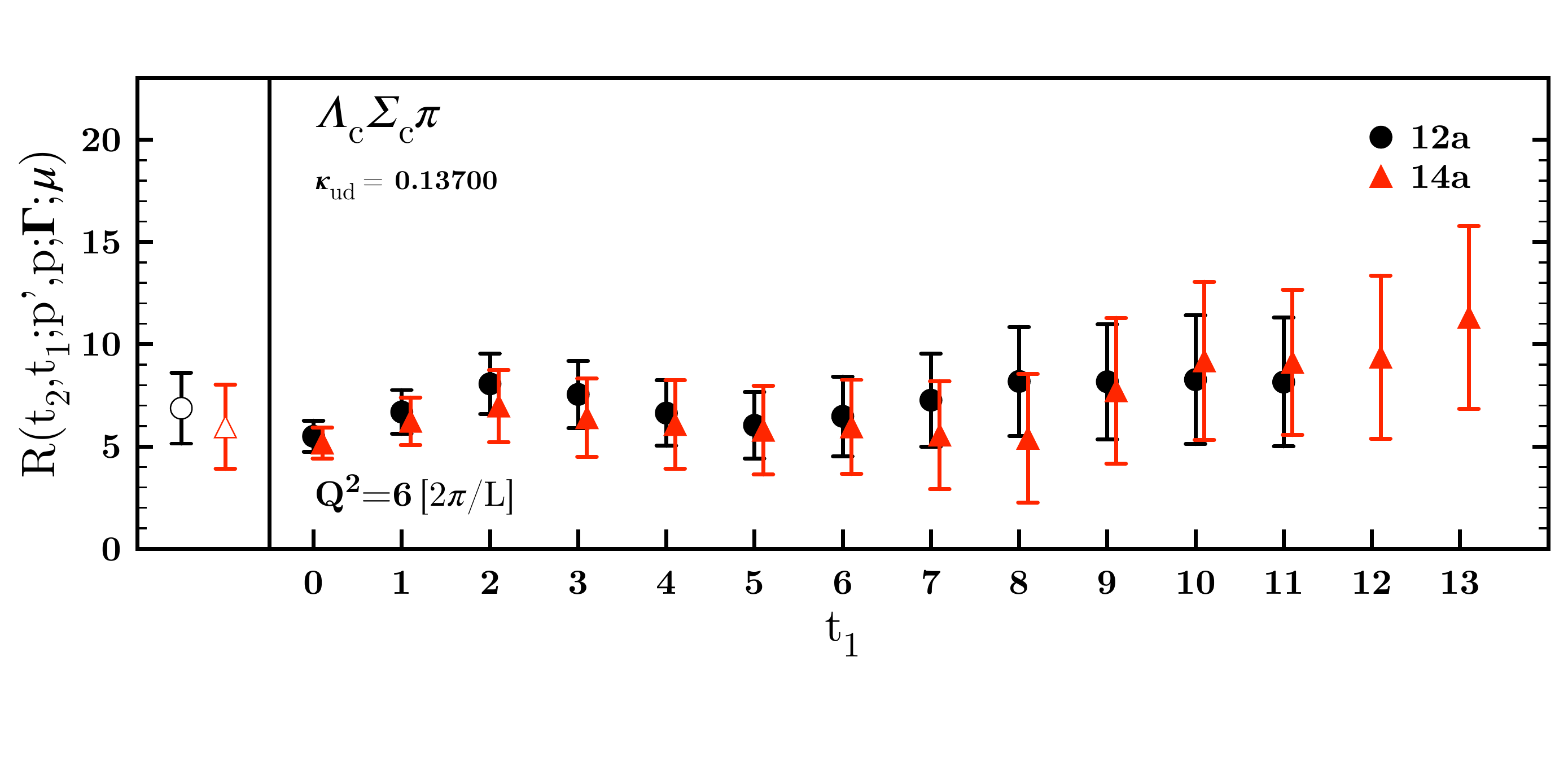}				
	\caption{A comparison of the behaviour of the Eq.~\ref{ratio} with respect to the current insertion time $t_1$ in case of two different source-sink seperations of  $12a$ and $14a$ for two different kinematical processes. Left panels hold the values extracted by a plateau analysis where the fit regions are chosen to be same for both seperations. $14a$ data points are shifted for clarity.}
	\label{fig:pl_1214}
\end{figure}
%%%%%%%%%%%%%%%%%%%%%%%%%%%%%%%%%%%%%%%%%%%%%%%%%

\begin{figure}[ht]
	\centering
	\includegraphics[width=.4\textwidth]{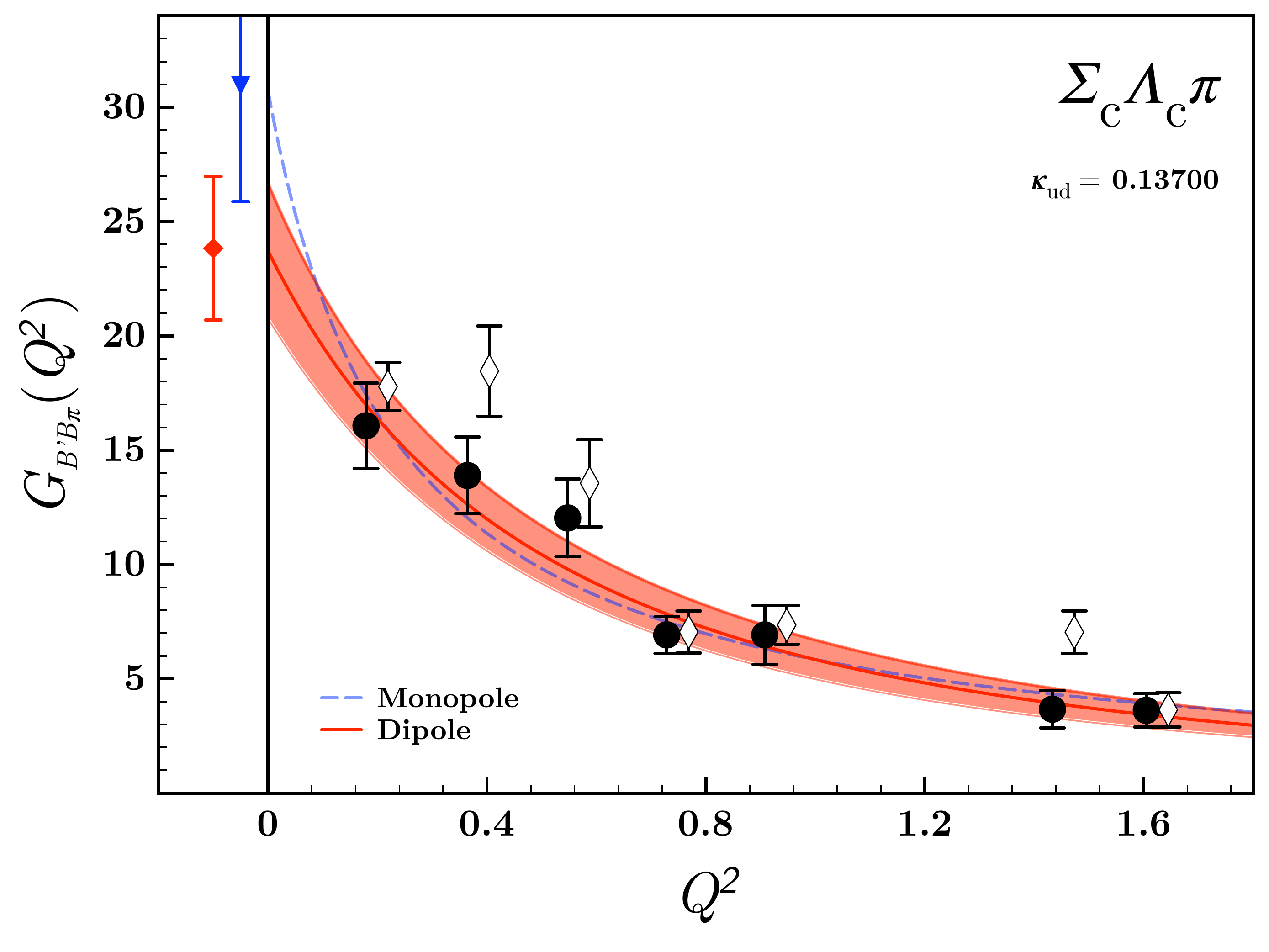} \includegraphics[width=.4\textwidth]{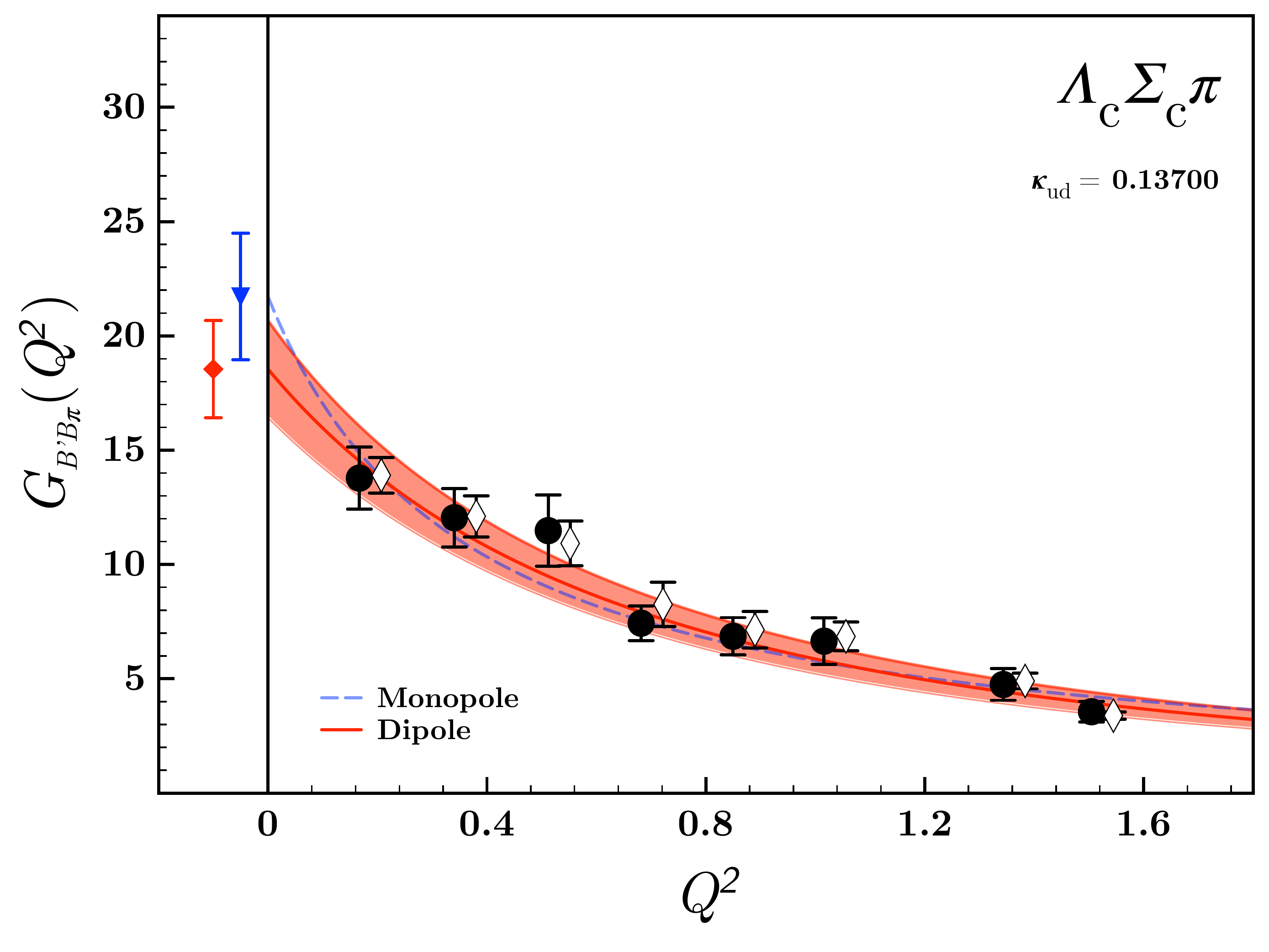} \\
	\includegraphics[width=.4\textwidth]{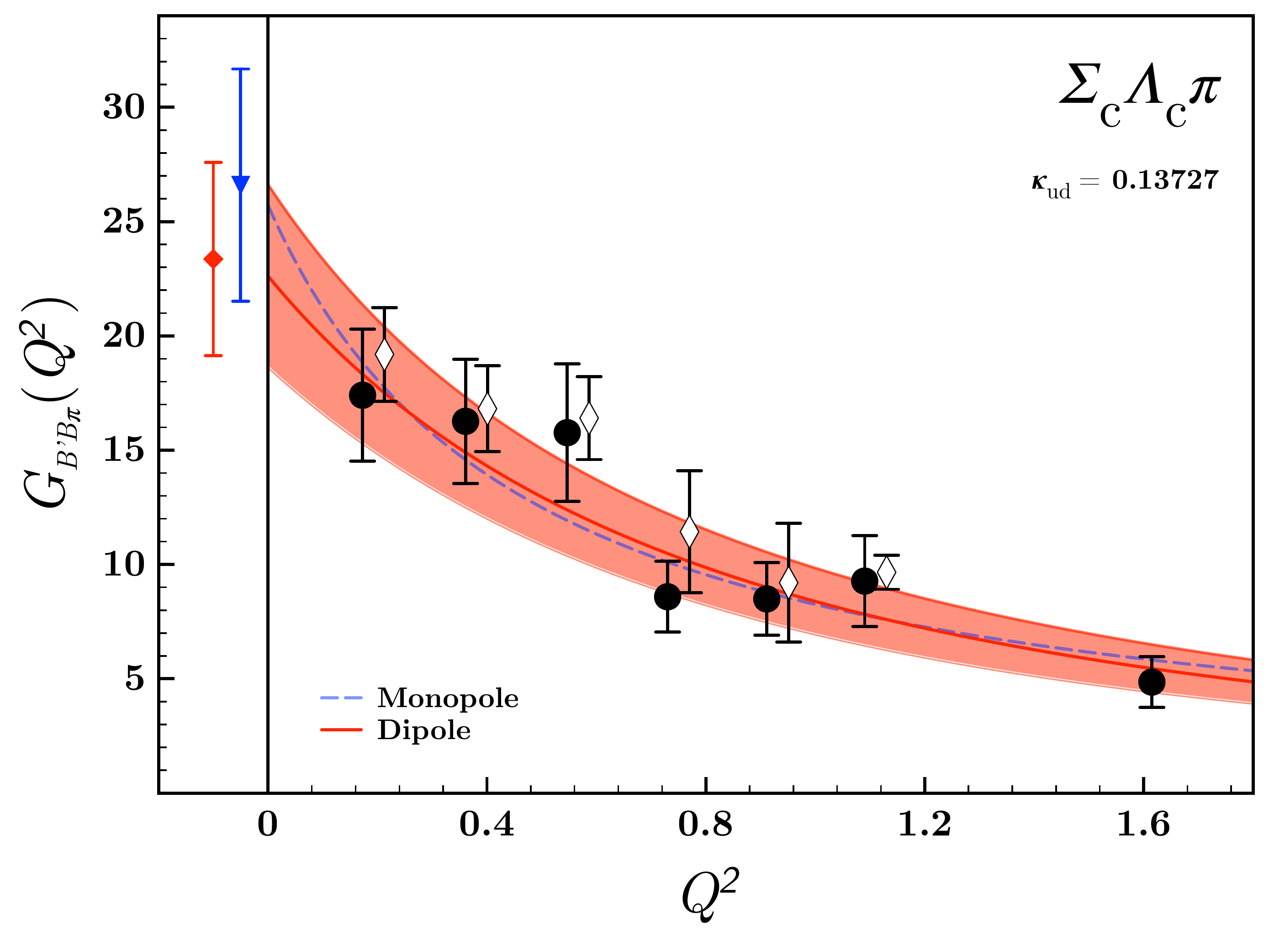} \includegraphics[width=.4\textwidth]{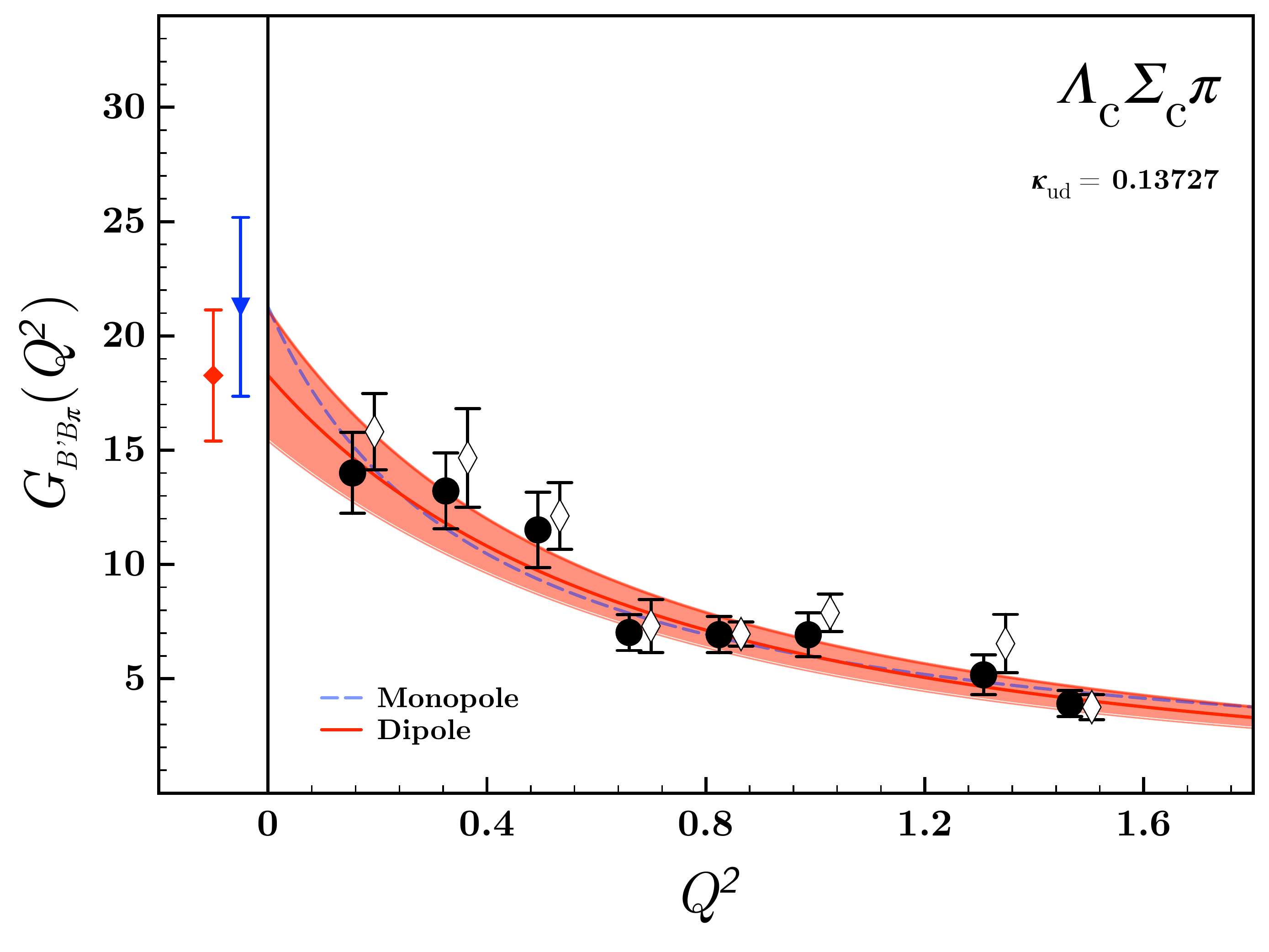} \\
	\includegraphics[width=.4\textwidth]{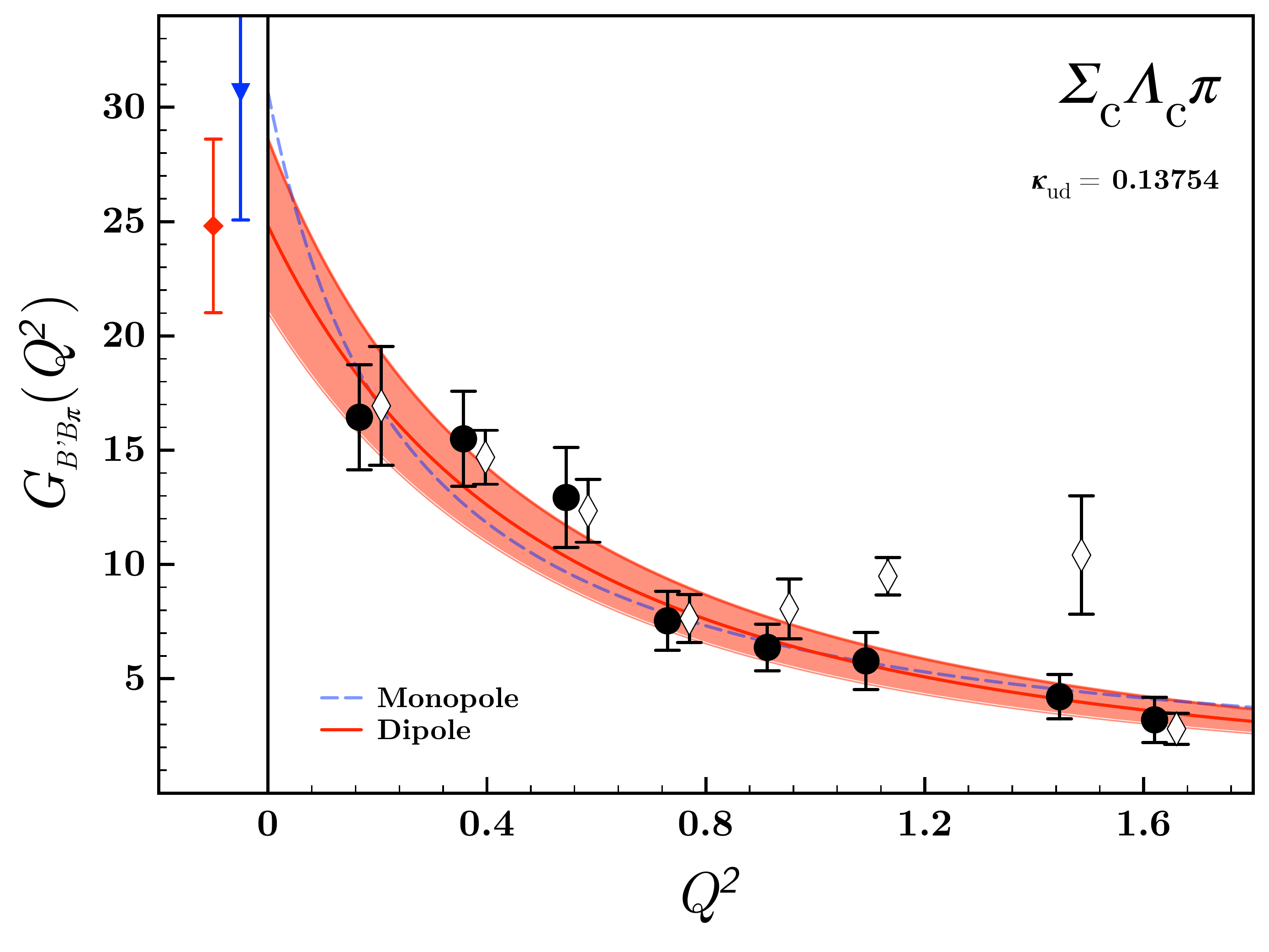} \includegraphics[width=.4\textwidth]{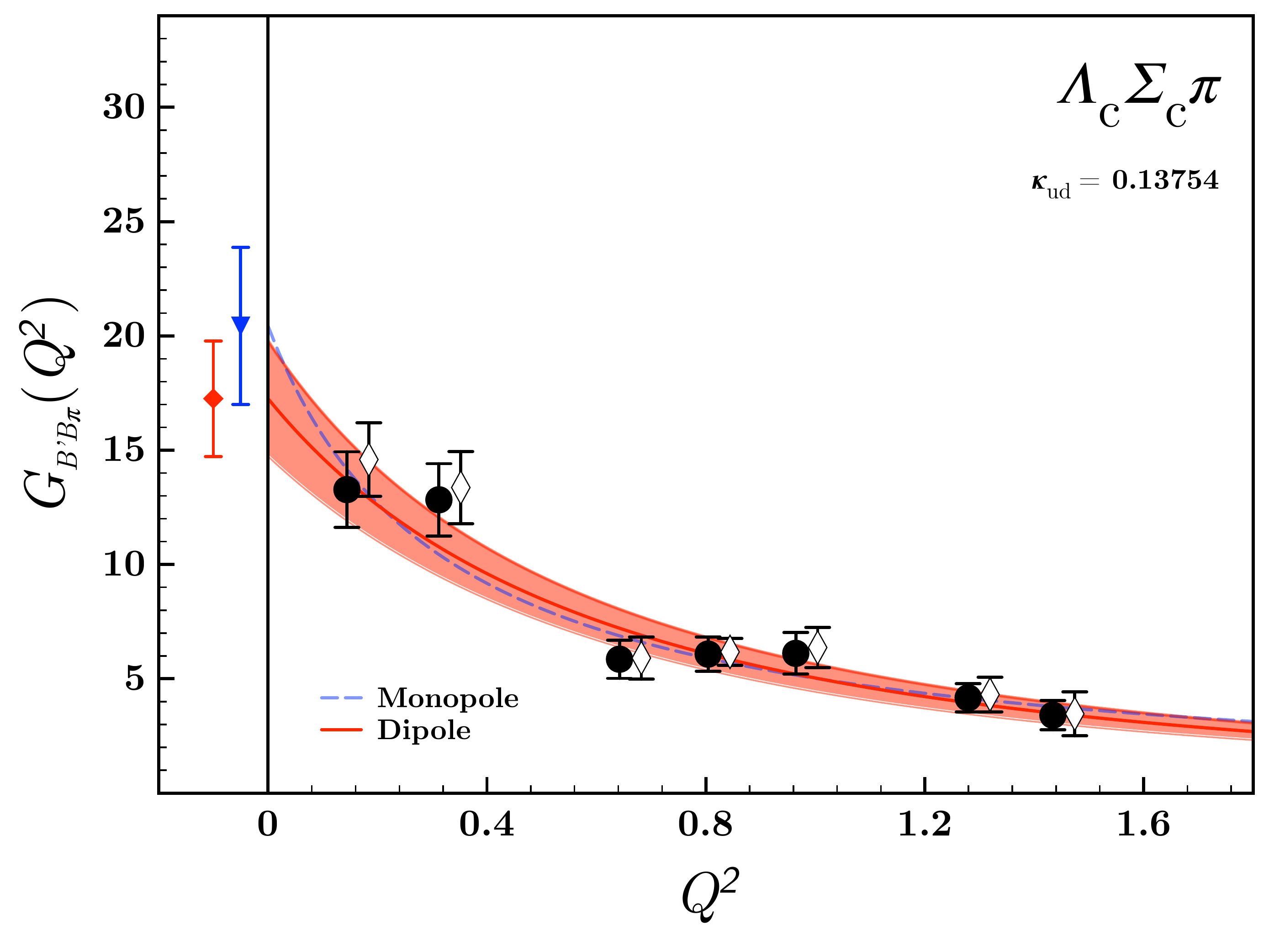} \\
	\includegraphics[width=.4\textwidth]{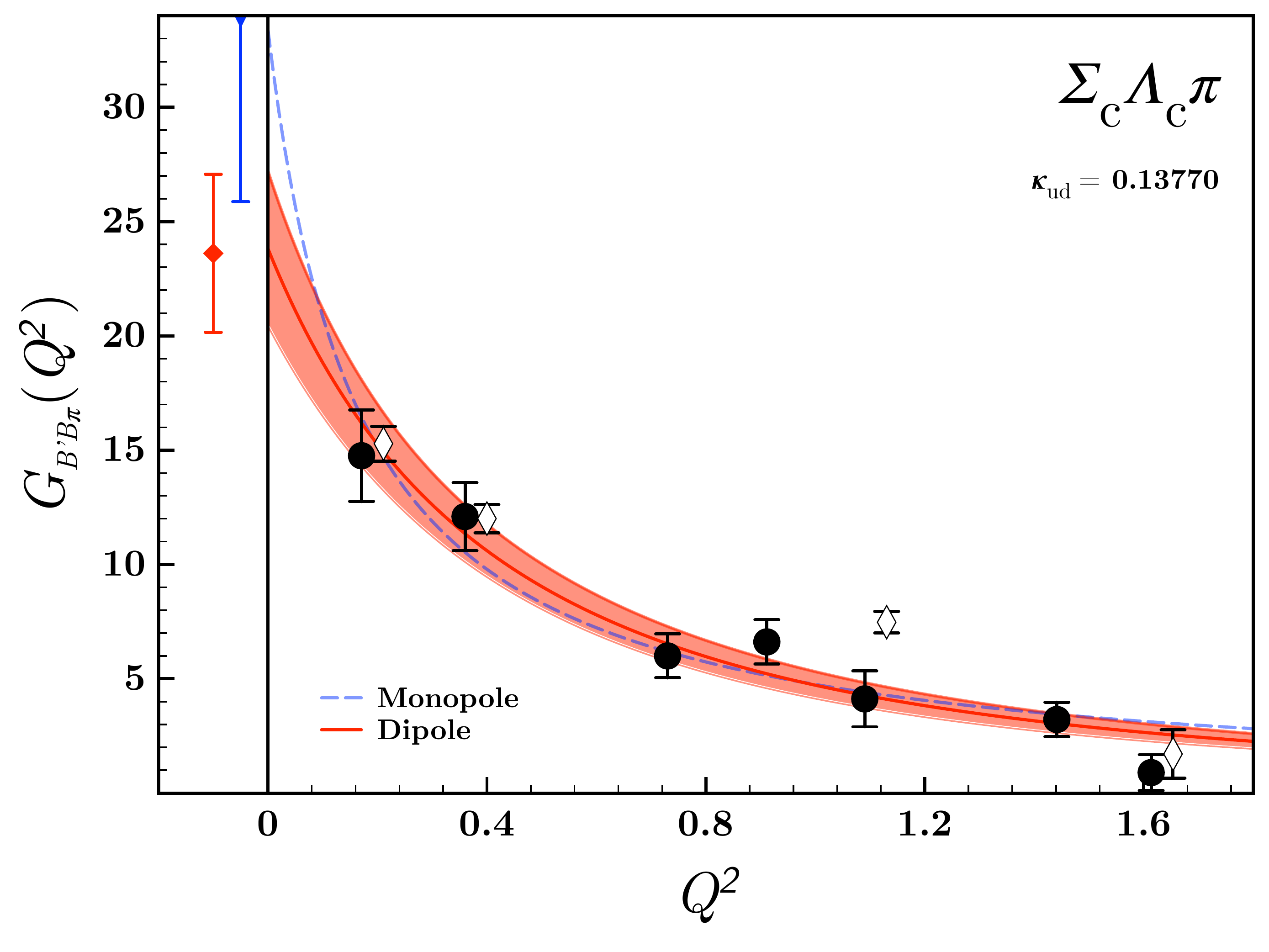} \includegraphics[width=.4\textwidth]{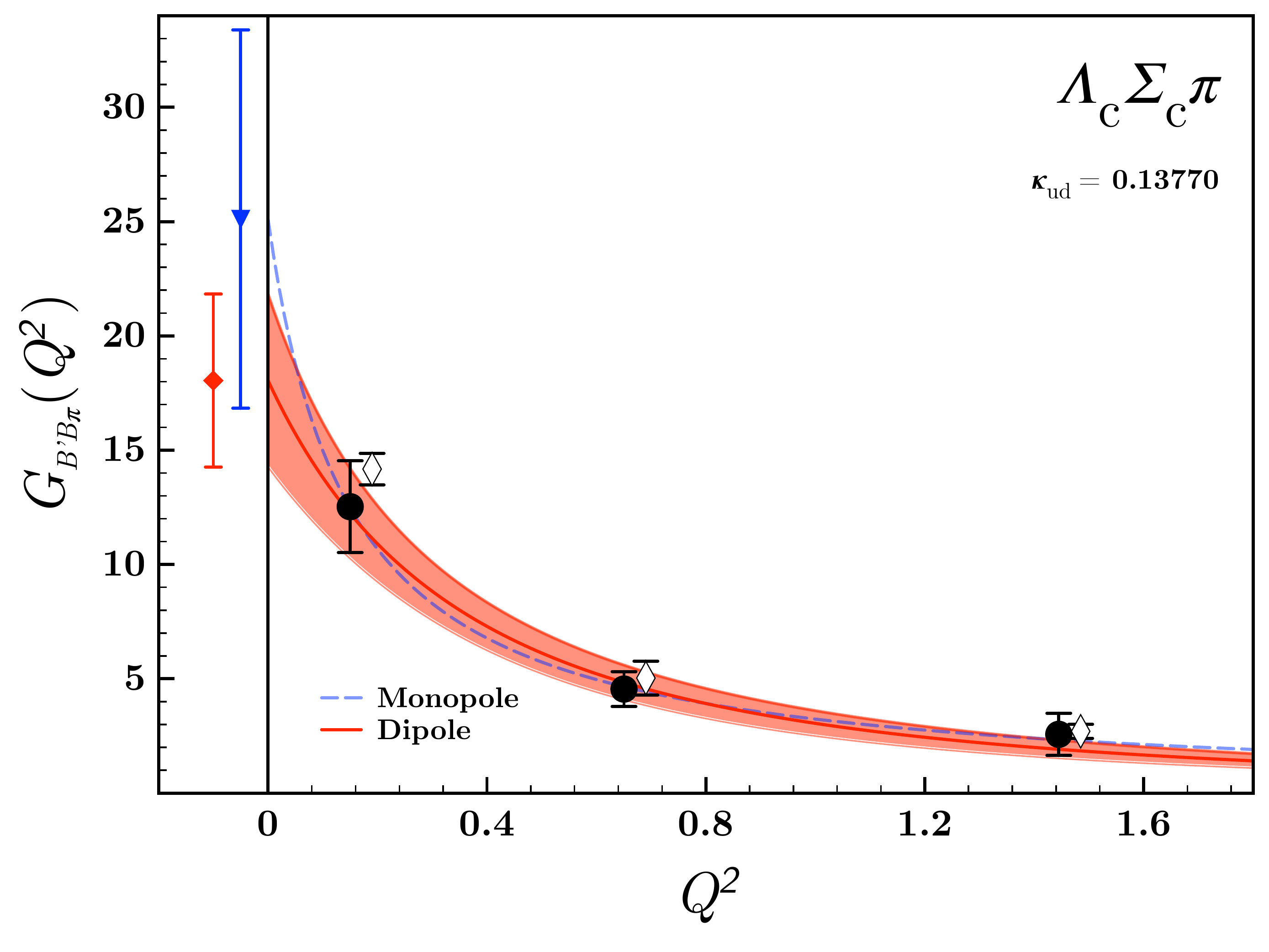}
	\caption{$\lsp$ (left) and $\slp$ (right) transition form factors computed on four different ensembles. Filled circles are values extracted by a plateau method whereas the empty diamonds are by the phenomenological form given in Eq.~\ref{epf}. We have omitted the values which have weak plateau signals. Lines of the best fit, error bands and the extrapolated values on the left panels are associated with plateau analysis.}
	\label{fig:ffs}
\end{figure}

Figure~\ref{fig:plepf} shows the ratio in Eq.~\ref{ratio} as a function of current-insertion time $t_1$ with $12a~(\sim 1.09~\rm{fm})$ seperation between the source and the sink on the heaviest quark ensemble ($\kappa^{u,d}=$0.13700) and for various momentum insertions. We compare the two form-factor values as extracted by the plateau method and by the phenomenological form fits. Apparent discrepancy between different fit procedures in the $\lsp$ kinematical case hints that either the data set is unreliable or the analysis suffers from excited-state contaminations. On the other hand, the $\slp$ case exhibits a good agreement between a plateau and a phenomenological approach. We observe a similar behaviour on the other ensembles also as shown in the Figure~\ref{fig:ffs}. We utilize the phenomenological form as a cross check rather than the actual fit procedure since regression analysis has a tendency to become unstable with increased number of free parameters. As long as the plateau fit results agree with that of the phenomenological form fit's we deem the data as reliable, less prone to excited state contamination and thus trust the identified plateaux and adopt its values for form factors.

As a further check of possible excited-state contaminations, we repeat the simulations on the $\kappa^{u,d}=$0.13700 ensemble with a larger source-sink seperation of $14$ lattice units $(\sim 1.27~\rm{fm})$. Figure~\ref{fig:pl_1214} shows the ratio in Eq.~\ref{ratio} as a function of current-insertion time for various momentum insertions with $t_2=12$ and $t_2=14$. In the case of $\lsp$ there is a large discrepancy between the $R(t_2,t_1;{\bf p^\prime},{\bf p};\Gamma;\mu)$ values of two different source-sink seperations and furthermore data are systematically smaller unlike the phenomenological form fit results. This inconsistency implies that not only the $\lsp$ case has significant excited state contamination but also the plateau and phenomenological-form fit analyses of the $12a$ data is unreliable. On the other hand, the $12a$ and $14a$ behaviour of the $\slp$ case is similar and consistent with the $12a$ phenomenological form analysis leading us to infer that $\slp$ is less affected by excited-state contaminations.

Figure~\ref{fig:ffs} illustrates the $\slpc$ and $\lspc$ form-factor measurements at eight momentum-transfer values available on the lattice. We show our results for all the ensembles $\kappa_{sea},\kappa_{val}^{u,d}=$ {0.13700, 0.13727, 0.13754, 0.13770}. While all form factors have a tendency to decrease as momentum transfer increases, there is a visible correlation amongst the data corresponding to first three and second three $Q^2$ values. Note that a similar behaviour also appears in the previous works on pseudoscalar-baryon coupling constants~\cite{Alexandrou:2007zz, Erkol:2008yj}. One possible source of this clustering with respect to momenta is the uncontrolled systematic errors such as discretization errors, which can be mitigated by use of finer lattices. In order to circumvent this problem one can analyze the on-axis (all momenta carried on a single axis; \emph{i.e.} ($p_x$,$p_y$,$p_z$)=(0,0,1), (0,0,2) and (0,0,3)) data only and perform a functional-form fit to extract the values at $Q^2=0$. Such an analysis however discards useful low-momentum data which is crucial to constrain the fits. We note that although we do not rely on this method, except in the $\kappa^{u,d}=$0.13770 case where the signal deteriorates heavily, our results given below differ by less than 3\% from those of an on-axis analysis.

We perform fits to $Q^2$ using pole-form ans\"{a}tze, \emph{viz.} a monopole form and a dipole form as given below, 
\begin{align}
	\label{ansatze}
		  G_{\mathcal{B}^\prime \mathcal{B} \pi}(Q^2) = \frac{G_{\mathcal{B}^\prime \mathcal{B} \pi}(0)}{1 + Q^2/\Lambda^2}, 
	\quad G_{\mathcal{B}^\prime \mathcal{B} \pi}(Q^2) = \frac{G_{\mathcal{B}^\prime \mathcal{B} \pi}(0)}{(1 + Q^2/\Lambda^2)^2},
\end{align}
where the $\Lambda$ is a free \emph{pole-mass} parameter. We require the extrapolated values to $Q^2=0$ using two ans\"{a}tze to be as close to each other as possible since the coupling constant value should be independent of the ansatz that's used to describe the form factors. We observe that such a condition is best realized in the $\slp$ case.
%%%%%%%%%%%%%%%%%%%%%% Figure - Form Factors %%%%%%%%%%%%%%%
% %%%%%%%%%%%%%%%%%%%%%%%%%%%%%%%%%%%%%%%%%%%%%%%

\begin{figure}[ht]
	\centering
	\includegraphics[width=.49\textwidth]{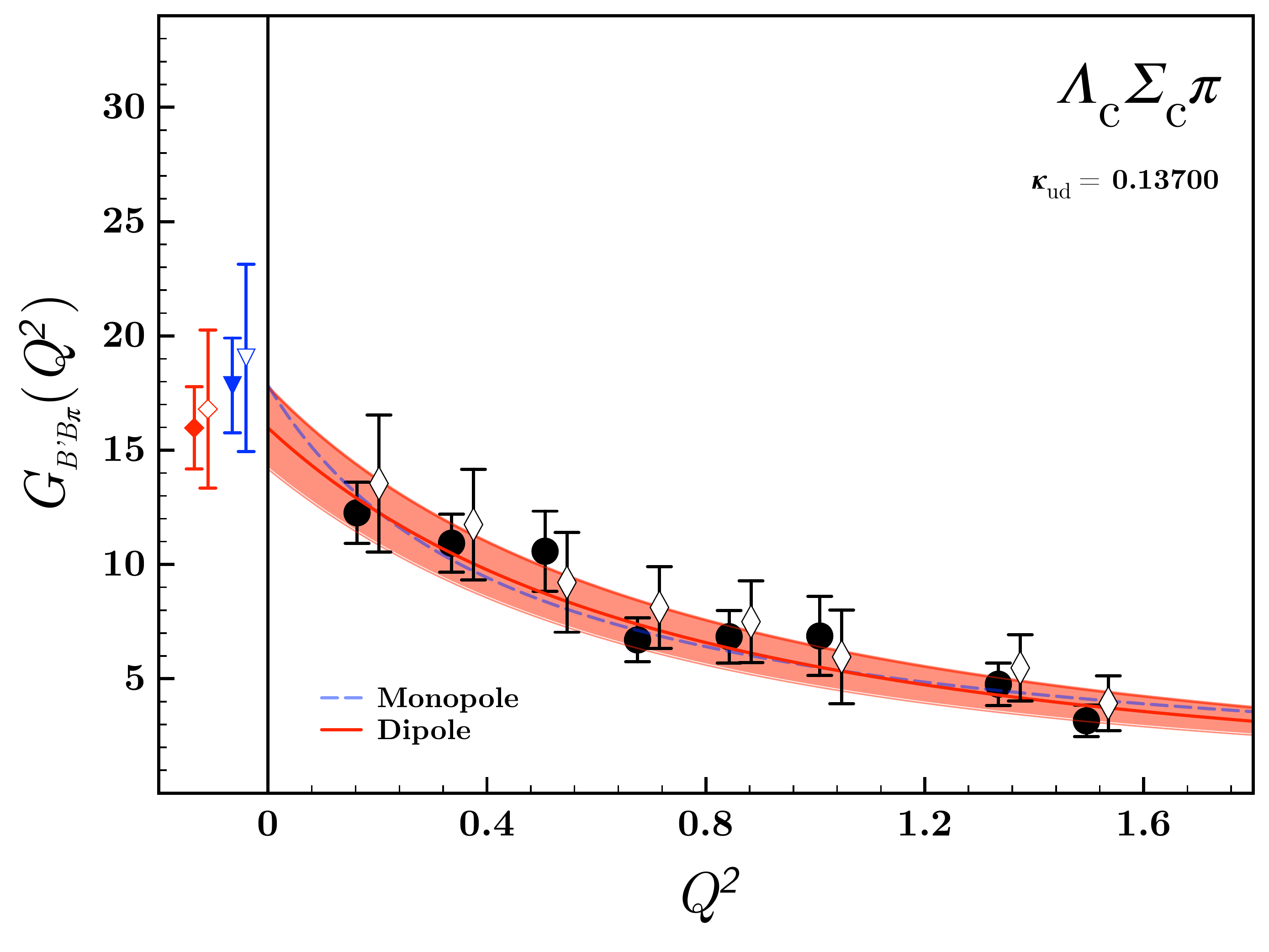} 
	\caption{$\slp$ transition form factor computed on $\kappa^{u,d}=$0.13700 ensemble.  Filled circles denote the $12a$ data where as the empty diamonds are $14a$ data. All the form factor values are extracted by the plateau analysis. Lines of the best fit and error bands are associated with $12a$ data. The extrapolated values on the left panels are for $12a$ (filled) and $14a$ (empty) data.}
	\label{fig:ff_1214}
\end{figure}

In order to make the final consideration to quantify the systematic errors arising due to the excited-state contamination, we visit the comparison of two cases with source-sink separation values once again and compare the extrapolated coupling constants. We show the plots of form factors with $t_2=12a$ and $t_2=14a$ in Figure~\ref{fig:ff_1214} where each data point is extracted by a plateau analysis. We focus particularly on the $\slp$ case for which the extrapolated values of the coupling constants by a dipole form are $G_{\lspc}^{12a} = 15.974(1.801)$ and $G_{\lspc}^{14a} = 16.797(3.462)$, where the discrepancy between the mean values is 5\%. Similarly, the final values of the coupling constants from a monopole fit differ by 7\% percent: $G_{\lspc}^{12a} = 17.835(2.071)$ and $G_{\lspc}^{14a} = 19.042(4.099)$. 

One important observation from the $\slp$ kinematical case in Figure \ref{fig:ff_1214} is that the correlation amongst the data mentioned above seems to vanish when the source sink separation is increased. However, any apparent correlation might be hidden by the increased statistical uncertainty. We have performed the $t_2=12a$ and $t_2=14a$ analysis with the same number of ensembles and the statistical errors increase roughly by 50\%. It would require at least twice as many measurements to reach a similar precision of $t_2=12a$ case. Although plausible for the $\kappa^{u,d}=$0.13700 case, this would not be possible for lighter quark-mass ensembles since the number of gauge configurations available is limited.       
%%%%%%%%%%%%%%%%% Figure - Form Factor 1214 %%%%%%%%%%%%%%%%%%%
% %%%%%%%%%%%%%%%%%%%%%%%%%%%%%%%%%%%%%%%%%%%%%%%%

Our conclusion from the above analysis is that the $\slp$ kinematical case with $t_2=12a$ source-sink separation is less prone to excited-state contaminations and therefore we give our final results considering the $\slp$ kinematical case only. We will assign a systematic error of minimum 6\% to the weighted averages of the coupling constants and propagate that error to the decay width in addition to the statistical errors. 

\begin{figure}[ht]
	\centering
	\includegraphics[width=.75\textwidth]{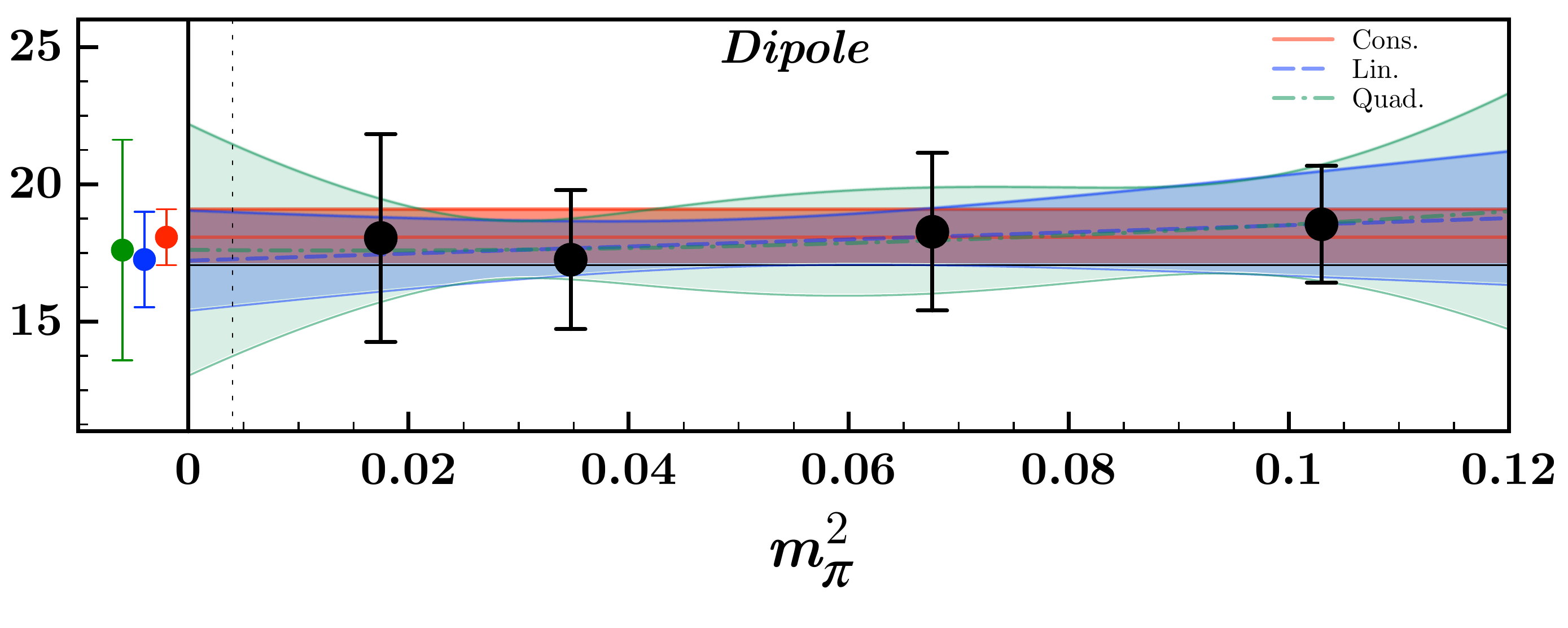} \\
	\includegraphics[width=.75\textwidth]{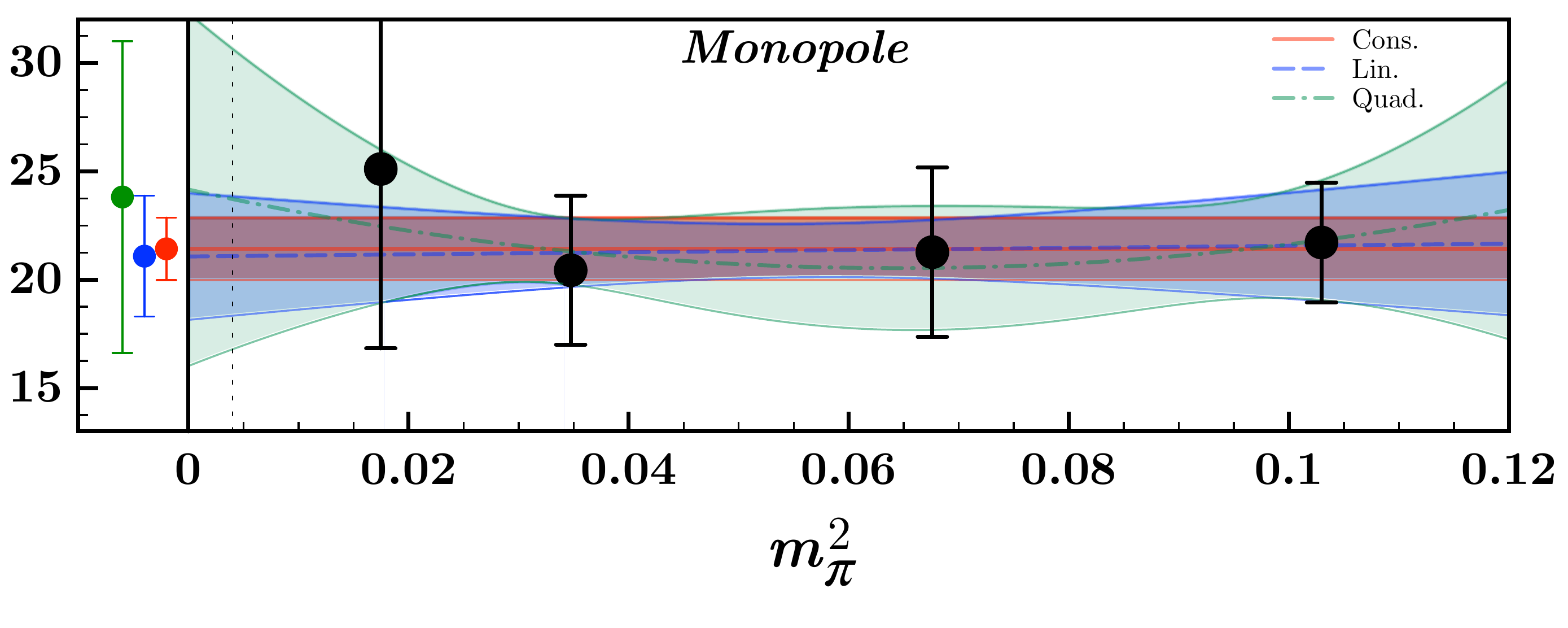}
	\caption{$G_{\lspc}$ coupling constant as a function of $m_\pi^2$ and extrapolation to the physical point. Points on the left panels are the extracted values on the physical quark-mass point indicated by a dashed vertical line.}
	\label{fig:chiralFits}
\end{figure}

We have tabulated the coupling constants as extracted on each ensemble with different functional forms in Table~\ref{res_table}. In Figure~\ref{fig:chiralFits} we show the $m_\pi^2$ dependence of the $G_{\lspc}(Q^2=0)$. We regard the deviation arising from different ans\"{a}tze used as a source of systematic error in our calculation and estimate the error by comparing the weighted average of monopole and dipole fit results to the dipole fit result on the physical point. Lower panel of Table~\ref{res_table} gives the results of the extrapolations to the physical point by a constant, by a linear and by a more general quadratic form in $m_\pi^2$. There is a reasonable agreement between the results of different extrapolation forms to the physical point. The weighted averages, reported on the final column of Table~\ref{res_table}, agree well with each other. 
%%%%%%%%%%%%%%%%%%%%% Figure - Chiral %%%%%%%%%%%%%%%%%%%
% %%%%%%%%%%%%%%%%%%%%%%%%%%%%%%%%%%%%%%%%%%%%%%
%%%%%%%%%%%%%%%%%%%%%%%%%%%% Table - Results %%%%%%%%%%%%%%%%%%%%%%%%%%%%%%
\begin{table*}[ht]
	\caption{Coupling constant values extracted on each ensemble by different ans\"{a}tze. Lower section contains the extrapolated values to the physical quark-mass point as well as the weighted averages. All results are also subject to at least 5\% excited state error in addition to the errors quoted in parentheses.}
	\label{res_table}
	\centering
	\setlength{\extrarowheight}{7pt}
	\begin{tabular*}{1.0\textwidth}{@{\extracolsep{\fill}}rcccc}
		\hline\hline
		\multirow{2}{*}{$\kappa^{u,d}_{val}$} & \multicolumn{2}{c}{$G_{\lspc}$} \\ 
		\cline{2-3}
				& Monopole Form & Dipole Form \\
		\hline \hline
		0.13700 	& 21.717(2.765) & 18.545(2.124) \\
		0.13727 	& 21.272(3.911) & 18.271(2.870) \\
		0.13754 	& 20.434(3.431) & 17.255(2.528) \\
		0.13770 	& 25.107(8.276) & 18.046(3.782) & $\bar{\mathit{x}}_w$($\hat{\sigma}_{\text{stat.}}$)($\hat{\sigma}_{\text{syst.}}$)\\
		\hline
		Const. Fit & 21.423(1.442) & 18.074(1.014) & 19.183(830)(1.109) \\
		Lin. Fit 	& 21.086(2.789) & 17.261(1.740) & 18.332(1.476)(1.071) \\
		Quad. Fit & 23.816(7.193) & 17.604(4.016) & 19.080(3.507)(1.476) \\
		\hline\hline
	\end{tabular*}
\end{table*}
%%%%%%%%%%%%%%%%%%%%%%%%%%%%%%%%%%%%%%%%%%%%%%%%%%%%%%%%%%%%%%%%%%

The final value we quote for the coupling constant is,
\begin{equation}
	G_{\lspc} = 18.332 \pm 1.476 \pm 2.171,
\end{equation}
where the first error is statistical and the second one is the combined systematical error due to weighted average and excited state contamination.

If we consider the decaying baryon at rest, the decay width of $\slp$ is given by \cite{Albertus:2005zy}
\begin{equation}
	\label{dw}
	\Gamma(\slp)=\frac{|\vec{q}_\pi|}{8\pi m_\Sigma^2} g^2_{\Lambda_c \Sigma_c \pi} ((m_\Sigma-m_\Lambda)^2-m_\pi^2),
\end{equation}
where $\vec{q}_\pi$ is the final pion three momentum in the rest frame of the decaying baryon
\begin{equation}
	\vec{q}_\pi=\frac{1}{2m_\Sigma}\lambda^{1/2}(m^2_\Sigma,m^2_\Lambda,m_\pi^2),
\end{equation}
with the Kallen function $\lambda(a,b,c)=a^2+b^2+c^2-2ab-2ac-2bc$. Using the physical values of the baryon masses reported by the PDG~\cite{Agashe:2014kda}, we evaluate the decay width given in Eq.\eqref{dw} as
\begin{equation}
	\Gamma_{\Sigma_c}=1.65 \pm 0.28_{\rm{stat.}} \pm 0.30_{\rm{syst.}}~\rm{MeV},
\end{equation}
which is in agreement with the recent experimental decay width determination of different isospin states as $\Gamma_{\Sigma_c^{++}}=1.84\pm 0.04^{+0.07}_{-0.20}$~MeV and as $\Gamma_{\Sigma_c^{0}}=1.76\pm 0.04^{+0.09}_{-0.21}$~MeV by Belle Collaboration~\cite{Lee:2014htd}. For comparison, we compile other theoretical determinations of the decay widths in the literature in Table~\ref{tab:comp}. In general other theoretical works tend to overestimate the $\Sigma_c$ decay width as compared to experiment and our lattice result. 
%%%%%%%%%%%%%%%%%%%%%%%%%%%%%%%%%%% Table - Comparison %%%%%%%%%%%%%%%%%%%%%%%%%%%%%%%%%%%%%
\begin{table*}[ht]
	\caption{ Comparison of our result with those from experiment~\cite{Lee:2014htd}, Heavy Hadron Chiral Perturbation Theory (HH$\chi$PT)~\cite{Cheng:2015naa, Huang:1995ke}, Light-front Quark Model (LFQM)~\cite{Tawfiq:1998nk}, Relativistic Quark Model (RQM)~\cite{Ivanov:1999bk}, Non-Relativistic Quark Model (NRQM)~\cite{Albertus:2005zy,Nagahiro:2016nsx}, $^3P_0$ Model \cite{Chen:2007xf} and QCD Sum Rules (QCDSR)~\cite{PhysRevD.79.056002} for the decay width of $\Sigma_c$. We quote either $\Gamma(\Sigma_c^{++}\rightarrow\Lambda_c \pi^+)$, $\Gamma(\Sigma_c^0\rightarrow\Lambda_c \pi^-)$ or the isospin average. All values are given in MeV.}
	\label{tab:comp}
	\centering
	\setlength{\extrarowheight}{5pt}
	\begin{tabular*}{1.0\textwidth}{@{\extracolsep{\fill}}cccccccccccccc}
		\hline\hline 
		& This Work 	& Experiment 			& HH$\chi$PT 			& HH$\chi$PT 			& LFQM 				& RQM 				& NRQM 			
& NRQM	& $^3P_0$ & QCDSR \\
		& 			& \cite{Lee:2014htd} 	& \cite{Huang:1995ke}		& \cite{Cheng:2015naa} 	& \cite{Tawfiq:1998nk} 	& \cite{Ivanov:1999bk} 	& \cite{Albertus:2005zy} & ~\cite{Nagahiro:2016nsx} & \cite{Chen:2007xf} & \cite{PhysRevD.79.056002} \\
		\hline \hline
		$\Gamma(\slp)$& $1.65(28)(30)$ & $1.80(4)^{+0.08}_{-0.21}$ & $2.5$ & $1.9^{+0.1}_{-0.2}$ & $1.48(17)$ & $2.75(19)$ & $2.39(7)$ & $4.27-4.33$ & $1.29$& $2.16(85)$ \\
		\hline\hline
	\end{tabular*}
\end{table*}
%%%%%%%%%%%%%%%%%%%%%%%%%%%%%%%%%%%%%%%%%%%%%%%%%%%%%%%%%%%%%%%%%%%%%%%%%%%%%%%%%%%

\section{Conclusion}
\label{sec:sum}
In summary, we have evaluated the $\Lambda_c\Sigma_c\pi$ coupling constant and the width of the strong decay $\Sigma_c \rightarrow\Lambda_c \pi$ in 2+1 flavor lattice QCD on four different ensembles with pion masses ranging from $\sim$ 700 to 300 MeV. A systematic analysis of different kinematical cases and the excited state contributions is given. Incorporating our results into the strong $\slp$ decay, we have obtained the decay width of $\Sigma_c$ as $\Gamma(\Sigma_c \rightarrow\Lambda_c \pi)=1.65(28)(30)$~MeV, which is in agreement with the experimental determination. 

\acknowledgments
This work is supported in part by The Scientific and Technological Research Council of Turkey (TUBITAK) under project number 114F261 and in part by KAKENHI under Contract Nos. 25247036, 24250294 and 16K05365. This work is also supported by the Research Abroad and Invitational Program for the Promotion of International Joint Research, Category (C) and the International Physics Leadership Program at Tokyo Tech.

\end{document}